\documentclass[usenatbib]{mn2e}

\usepackage{amsmath,amssymb}
\usepackage{graphicx}
\usepackage[usenames,dvipsnames]{color}
\usepackage{multirow}

\newcommand{\mAa}{A1}
\newcommand{\mAb}{A2}
\newcommand{\mAc}{A3}
\newcommand{\mD}{B1}
\newcommand{\mE}{B2}
\newcommand{\mF}{C1}
\newcommand{\mG}{C2}
\newcommand{\mH}{C3}
\newcommand{\mI}{C4}
\newcommand{\mJ}{C5}
\newcommand{\mK}{D1}
\newcommand{\mL}{D2}
\newcommand{\mM}{D3}
\newcommand{\mN}{D4}
\newcommand{\mO}{D5}
\newcommand{\mEa}{E1}
\newcommand{\mEb}{E2}
\newcommand{\mEc}{E3}
\newcommand{\mEd}{E4}
\newcommand{\mEe}{E5}
\newcommand{\mP}{F1}
\newcommand{\mQ}{F2}
\newcommand{\mR}{F3}
\newcommand{\mGa}{G1}
\newcommand{\mGb}{G2}
\newcommand{\mHa}{H1}
\newcommand{\mHb}{H2}
\newcommand{\mHc}{H3}
\newcommand{\mHd}{H4}
\newcommand{\mIa}{I}

\title{Shocks in the relativistic transonic accretion with low angular momentum}

\author[P. Sukov\'a et al.]
       {P. Sukov\'a, $^{1}$\thanks{E-mail: petra.sukova@asu.cas.cz}
        S. Charzy\'nski, $^{2,3}$
        and
       A. Janiuk  $^{2}$ \thanks{E-mail: agnes@cft.edu.pl}
       \\
       $^{1}$Astronomical Institute, Czech Academy of Sciences, Bo\v{c}n\'i II 1401, 141 00 Prague, Czech Republic\\
       $^{2}$Center for Theoretical Physics PAS, Al. Lotnik\'ow 32/46, 02-093 Warszawa, Poland\\
       $^{3}$Chair of Mathematical Methods in Physics, University of Warsaw, ul. Pasteura 5, 02-093 Warszawa, Poland
       }

\begin{document}

\date{}

\pagerange{\pageref{firstpage}--\pageref{lastpage}} \pubyear{}

\maketitle

\label{firstpage}

\begin{abstract}
We perform 1D/2D/3D relativistic hydrodynamical simulations of accretion flows with low angular momentum, filling the gap between spherically symmetric Bondi accretion and disc-like accretion flows. Scenarios with different directional distributions of angular momentum of falling matter and varying values of key parameters such as spin of central black hole, energy and angular momentum of matter are considered. In some of the scenarios the shock front is formed. We identify ranges of parameters for which the shock after formation moves towards or outwards the central black hole or the long lasting oscillating shock is observed. The frequencies of oscillations of shock positions which can cause flaring in mass accretion rate are extracted. The results are scalable with mass of central black hole and can be compared to the quasi-periodic oscillations of selected microquasars (such as GRS 1915+105, XTE J1550-564 or IGR J17091-3624), as well as to the supermassive black holes in the centres of weakly active galaxies, such as Sgr $A^{*}$.
\end{abstract}


\begin{keywords}
accretion, accretion discs -- hydrodynamics -- shock waves -- X-rays: binaries -- stars: black holes -- Galaxy: centre
\end{keywords}



\section{Introduction} \label{s:Introduction}

The weakly active galaxies, where the central nucleus emits the radiation at a moderate
level with respect to the most luminous quasars, are frequenlty described by the so called hot accretion flows model. In such flows, the plasma is virially hot and optically thin, and due to the advection of energy onto black hole, the flow is radiativelly inefficient. The prototype of an object which
can be well described by such model, is the low luminosity black hole in the centre of our Galaxy,
the source Sgr $A^{*}$ \citep{2014ARA&A..52..529Y}. Also, the black hole X-ray binaries in their hard and
quiescent states can be good representatives for the hot mode of accretion. These states are
frequenlty associated with the ejection of relativistic streams of plasma, which form the jets, responsible for the radio emission of the sources \citep{2004MNRAS.355.1105F}.

The matter essentially flows into black hole with the speed of light,
while the sound speed at maximum can reach $c/\sqrt{3}$. Therefore, the
accretion flow must have transonic nature. The viscous accretion
with transonic solution based on the alpha-disk model was first studied by
\cite{1981AcA....31..283P} and \cite{1982AcA....32....1M}.
After that, e.g. \cite{1989ApJ...336..304A} examined the stability and
structure of transonic disks.
The possibility of collimation of jets by thick accretion tori was proposed by, e.g., \cite{1981MNRAS.197..529S}.

In respect of the value of angular momentum there are two main regimes of accretion, the Bondi accretion, which refers to spherical accretion of gas without any angular momentum, and the disk-like accretion with Keplerian distribution of angular momentum. In the case of the former, the sonic point is located farther away from the compact object (depending on the energy of the flow) and the flow is supersonic downstream of it. In the latter case, the flow becomes supersonic quite  close to the compact object. For gas with a low value of constant angular momentum, hence belonging in between these two regimes, the equations allow for the existence of two sonic points of both types.

The possible existence of shocks in low angular momentum flows connected with the presence of multiple critical points in the phase space has been studied from different points of view during the last thirty years.

 Quasi-spherical distribution of the gas endowed by constant specific angular momentum $\lambda$ and the arisen bistability was studied already by~\cite{1981ApJ...246..314A}.
 \cite{1987PASJ...39..309F} studied the existence of critical points for realistic equation of state of the gas and showed the corresponding Rankine-Hugoniot conditions for standing shocks.
Later, the significance of this phenomenon related to the variability of some X-ray sources has been pointed out by \cite{1988ChA&A..12..119J} and soon after, the possibility of the shock existence together with shock conditions in different types of geometries was discussed also by \cite{1990ApJ...350..281A}.

The transonic solution required by the aforementioned boundary conditions is the solution  having low subsonic velocity far away from the compact object ($\mathfrak{M}<1$, where $\mathfrak{M}=v/c_{\rm s}$ is the Mach number, $v=-u^r_{\rm BL}$ is the inward radial velocity in Boyer-Lindquist coordinates, and $c_{\rm s}$ is the local sound speed of the gas), and supersonic velocity very near to the horizon ($\mathfrak{M}>1$). Thus the flow can locally pass only through the outer or also through the inner sonic point. The latter is globally achieved due to the shock formation between the two critical points.

More recently, the shock existence was found also in the disc-like structure with low angular momentum in hydrostatic equilibrium  both in pseudo-Newtonian potential \citep{2002ApJ...577..880D} and in full relativistic approach \citep{2012NewA...17..254D}.
Regarding the sequence of steady solutions with different values of specific angular momentum, the hysteresis-like behavior of the shock front was proposed in the latter work.
Different geometrical configurations with polytropic or isothermal equation of state were studied in the post Newtonian approach with pseudo-Kerr potential \citep{SAHA201610}.
In the general relativistic description the dependence of the flow properties (Mach number, density, temperature and pressure) in the close vicinity of the horizon was studied by \cite{Das201581}, and the asymmetry of prograde versus retrograde accretion was shown.

The complete picture of the accreting microquasar consisting of the Keplerian cold disc and the low angular momentum hot corona, the so called two component advective flow (TCAF), was described by \cite{1995ApJ...455..623C}. This model combined with the propagating oscillatory shock (POS) model was later used to explain the evolution of the low frequency QPO during the outburst in several microquasars \citep{2008A&A...489L..41C,2009MNRAS.394.1463C,2012A&A...542A..56N,2013arXiv1306.3745D} and it was also implemented into the \texttt{XSPEC} software package \citep{doi:10.1093/mnrasl/slu024}.

The presence of the low angular momentum component in the accretion flow during the outbursts of microquasars seems to be essential for explaining different timing in the hard and soft bands, especially during the rising phase \citep{0004-637X-565-2-1161}. \cite{DEBNATH20132143} showed with the spectral fitting, that the flux of the power-law component is higher than the flux of the disc black body in the time period, when the QPOs are seen in H~1743-322. \cite{doi:10.1093/mnrasl/slu024} showed similar results with the TCAF model, which gives the mass accretion rate of the disc component comparable to the accretion rate of the sub-Keplerian component, when the QPO is seen in three different sources. 

Further development in this topic includes numerical simulations of low angular momentum flows in different kinds of geometrical setup. Hydrodynamical models of the low angular momentum accretion flows have been studied already in two and three
dimensions, e.g.~by \cite{2003ApJ...582...69P}, \cite{2008ApJ...681...58J} and \cite{2009ApJ...705.1503J}. In those simulations,
a single, constant value of the specific angular momentum was assumed, while the variability of the flows
occurred due to e.g.~non-spherical or non-axisymmetric distribution of the matter.
The level of this variability
was also dependent on the adiabatic index.
However, these studies have not concentrated on the existence of the standing shocks as predicted by the theoretical works mentioned above.

The consideration was also put on the problem of viscosity in such flows, especially on the influence of the viscosity on the position of the shock and the shape of the solution \citep{1990MNRAS.243..610C,2004MNRAS.349..649C,2016ApJ...816....7N}. The possible consequences of viscous mechanisms in the shocked accretion flow for the QPOs' evolution was studied by \cite{0004-637X-798-1-57}. Later on, \cite{2016MNRAS.462..850N} added the phenomenon of outflows to the picture.
\cite{doi:10.1093/mnras/stw1327} showed the joint role of viscosity and magnetic field considered in heating of the accretion flow combined with the cooling by Comptonization on the dynamical structure of the global accretion flow.
Such models were also studied by hydrodynamical numerical simulations in the pseudo-Newtonian description of gravity e.g. by \cite{2010MNRAS.403..516G,2013MNRAS.430.2836G,2016MNRAS.462.3502D}.

Our aim is to provide numerical simulations of low angular momentum flows, which would support or correct the semi-analytical findings about the shock existence and behavior mentioned above. In our previous work, we performed 1D pseudo-Newtonian computations \citep{our_paper}, where we studied the dependence of the shock solution of the parameters and the response of the shock front to the change of angular momentum in the incoming matter. The hysteresis behavior was observed in our simulations and we have seen the repeated creation and disappearance of the shock front due to the oscillations of angular momentum in the flow.
Here we aim to provide more advanced numerical study of the flow using the full relativistic treatment of the gravity with the fixed background metric given by the Schwarzschild/Kerr solution, which is performed in one, two and three dimensions.

The organization of the paper is as follows. 
In Section \ref{s:Shocks} we briefly recall the semi-analytical treatment of the shock existence in the pseudo-Newtonian approach, which is described in details in \cite{our_paper}. The numerical setup of our simulations is given in Section \ref{s:Numerics}, the different initial conditions are described in Subsections \ref{Ini_Bondi}, \ref{Ini_shock} and \ref{Ini_spin}. In Section \ref{s:Results} we present our results, in particular in \ref{results_1D} we show the 1D simulations with standing or oscillating shock location and we run models with time-dependent outer boundary condition corresponding to periodic change of angular momentum in the incoming matter. The major part of the results is presented in \ref{results_2D}, where the outcomes of the 2D simulations with different kind of initial conditions are presented. In \ref{3D} we confirm the reliability of the 2D results with two full three dimensional tests. The findings of our study are discussed in Section \ref{s:Conclusions}.

\section{Appearance of shocks in 1D low angular momentum flows}\label{s:Shocks}

In this paper we follow up our previous study of the flow with constant low angular momentum $\lambda$, which was held in the pseudo-Newtonian framework.
Here we  briefly recall the semi-analytical results in such setup, which we use as an initial setting for our GR computations (for further details see \citet{our_paper}).
 For the analytical study we consider a non-viscous quasi-spherical polytropic flow with the equation of state $p=K \rho^\gamma$, where $p$ is the pressure, and $\rho$ is the density in the gas.
Our EOS holds for the isentropic flow, hence the specific entropy $K$ is constant.

Using the continuity equation and energy conservation, we can find the position of the critical point $r_c$ as the root of the equation
\begin{multline}
\mathcal{E} - \frac{\lambda^2}{2r_c^2}  + \frac{1}{2(r_c-1)}-\\
\frac{\gamma+1}{2(\gamma-1)}  \left( \frac{r_c}{4(r_c-1)^2} - \frac{\lambda^2}{2r_c^2} \right) = 0, \label{r_c}
\end{multline}
where $\mathcal{E}$ stands for energy and where we assume the Paczynski-Wiita gravitational potential in the form of  $\Phi(r)=-\frac{1}{2(r-1)}$, so that $r$ is given in the units of $r_g=2GM/c^2$, and where $\lambda$ is the value of specific angular momentum.

For a subset of the parameter space ($\mathcal{E}, \lambda, \gamma$) there exists more than one solution of this equation (three actually), hence there are more critical points located at different positions. It can be shown, that only two of the critical points are of a saddle type, so that the solution can pass through it. We will call them the inner, $r_c^{\tt in}$, and the outer, $r_c^{\tt out}>r_c^{\tt in}$, critical points, respectively. We will call this subset as ``multicritical region''. For changing $\lambda$ with other parameters kept constant, this region is projected onto one interval of $[\lambda_{\tt min}^{cr},\lambda_{\tt max}^{cr}]$. For decreasing $\mathcal{E}$, the interval is shifting up to a higher angular momentum.

Together with equation (\ref{r_c})  determining the values of all variables at the two critical points, the relation for the derivative ${\rm d}v/{\rm d}r$ can be obtained from the continuity equation and the energy conservation, so that the solution can be found by integrating the equations from the critical point downwards and upwards. The resulting two branches of solutions of course have the same parameters ($\mathcal{E}, \lambda, \gamma$), but they differ in value of the constant specific entropy $K$, which is given by
\begin{equation}
K = \left( v r^2 \frac{c_s^{\frac{2}{\gamma - 1}}}{\gamma^{\frac{1}{\gamma-1}} \dot{M}} \right)^{\gamma - 1}. \label{konst_K}
\end{equation}
This is evaluated at the critical point position ($K^{\rm in/out}=K(r^{\rm in/out}, v^{\rm in/out},c_s^{\rm in/out})$, where $\dot{M}$ is the adopted constant accretion rate. Because in our model we study the motion of test matter, which does not contribute to the gravitational field, and we use a simple polytropic equation of state, the accretion rate can be given in arbitrary units and it does not affect the solution.

The only possible production of entropy is at the shock front, where jumps in radial velocity, density, and other quantities in the flow occur. Because we are interested in the solution describing the accretion flow, and not winds, the physical requirement for the shock to occur is that the specific entropy at the inner branch is higher than at the outer one ($K^{\tt in} > K^{\tt out}$). Moreover, the shock will appear only if the Rankine-Hugoniot conditions, expressing the conservation of mass, energy and angular momentum at the shock position are satisfied, that means that the equation
\begin{equation}
\frac{ \left(\frac{1}{\mathfrak{M}_{\tt in}}  + \gamma \mathfrak{M}_{\tt in} \right)^2 }{ \mathfrak{M}_{\tt in}^2(\gamma - 1) + 2 } = \frac{ \left(\frac{1}{\mathfrak{M}_{\tt out}}  + \gamma \mathfrak{M}_{\tt out}\right)^2 }{ \mathfrak{M}_{\tt out}^2(\gamma - 1) + 2 } \label{shock}
\end{equation}
holds at some radius $r_s$.

\section{Numerical setup} \label{s:Numerics}

We performed 1D to 3D hydrodynamical simulations of the non-magnetized accreting gas on the fixed background using the \texttt{HARMPI} \footnote{e.g., see \texttt{https://github.com/atchekho/harmpi}} computational code \citep{2015MNRAS.454.1848R,2017MNRAS.467.3604R,2012MNRAS.423.3083M,2007MNRAS.379..469T,2007MNRAS.378.1118M, 2013ApJ...776..105J, 2017ApJ...837...39J} based on the original HARM code \citep{0004-637X-589-1-444,0004-637X-611-2-977}.
The background spacetime is given by the stationary Kerr solution.

The initial conditions are set using Boyer-Lindquist coordinates, and they are transformed into the code coordinates, which are the Kerr-Shild ones.
In order to cover the whole accretion structure with sufficiently fine resolution near the black hole we use logarithmic grid in radius with superexponential grid spacing in the outermost region, so that the outer region is covered with low resolution grid and provide the reservoir of gas for accretion.
In the innermost region, the grid spans below the horizon thanks to the regularity of Kerr-Shild coordinates, having several zones inside the black hole and the free outflow boundary.
The outer boundary condition is mostly given as a free boundary, because the outer boundary is placed far enough from the central region. In the 1D case, when long-term evolution is studied ( $t_f$ up to $5\cdot 10^7 {\rm M}$)), we prescribe the inflow of matter through the outer boundary according to the PN solution. Because the outer boundary is very far away from the centre (typically at $\sim 10^5$~M), the radial inflowing speed is very small and the deviations between GR and PN solution are negligible.
The prescribed properties of the inflowing matter can be also time dependent, when the temporal change of the angular momentum of the matter is considered.
For 2D computations, the resolution is in the range between 256 x 128, up to 384 x 256 and 576 x 192, while in 3D case we use 256 x 128 x 96 cells.

\begin{figure}
 \includegraphics[width=0.48\textwidth]{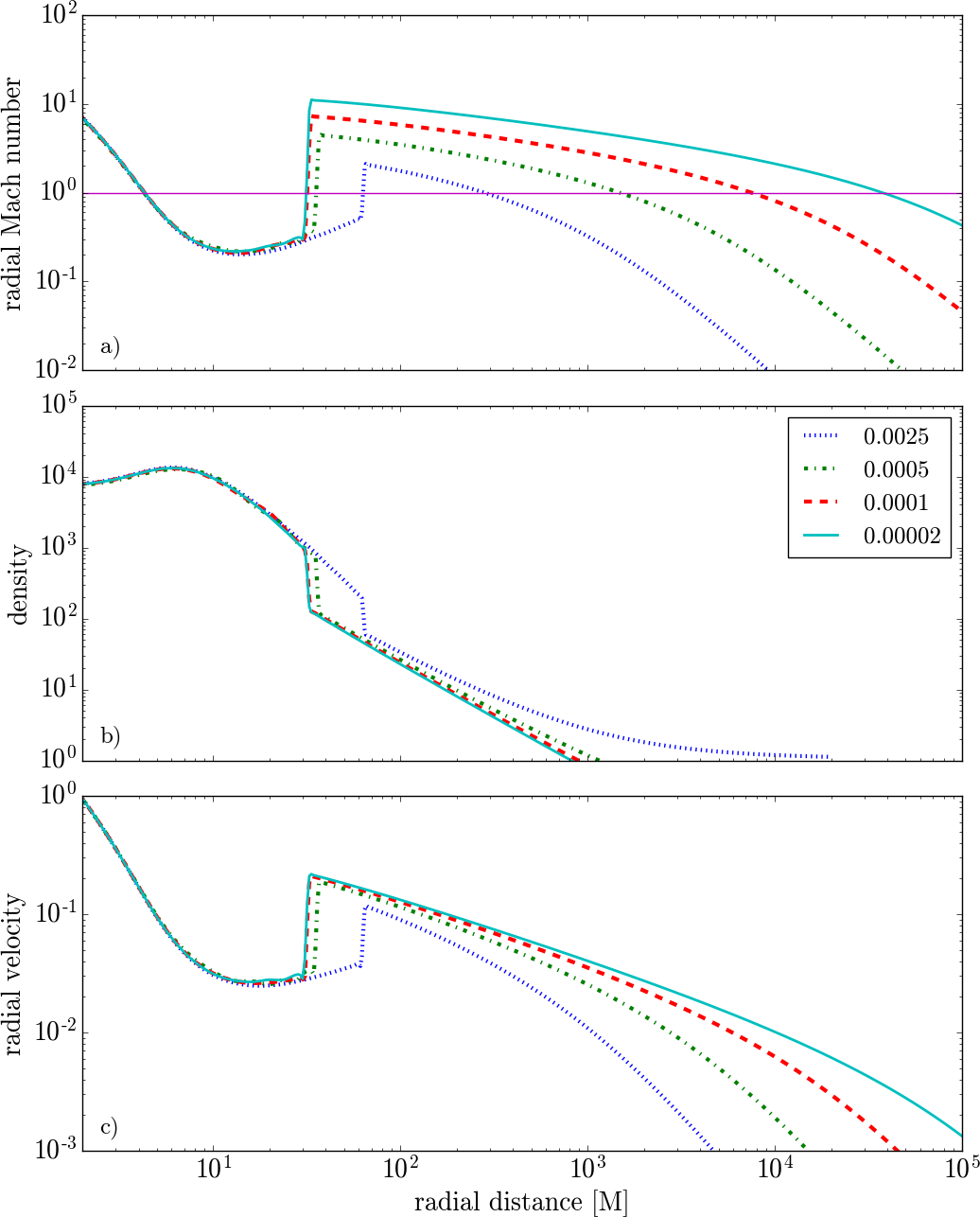}
   \caption{a) Profiles of Mach number for four values of energy in the converged stationary state for $\gamma=4/3$ and $\lambda = 3.6M$ at the end of the simulation, $t_f = 10^6$M. Sonic points and shock fronts are located at the points, where the curve crosses the $\mathfrak{M}=1$ line (purple horizontal line). b) The corresponding profiles of density in arbitrary units. c) Radial velocity profiles of the flow $v=- u_{\rm BL}^{\tt r}$ in the units of the speed of light, $c$. \label{1D_sol}}
\end{figure}

In our previous work, \cite{our_paper}, we studied the behavior of the shock solution and also its evolution with 1D simulations using the code \texttt{ZEUS-MP} \citep{1992ApJS...80..753S,2003ApJS..147..197H} supplied by the pseudo-Newtonian Paczynski-Wiita potential \citep{1980A&A....88...23P}, which mimics the strong gravity effects near the black hole.
We will refer to the results presented in that paper as the PW simulations.
The parameters which we used in that work, corresponded either to the evolving frequency of quasi-periodic oscillations seen in some  microquasars (e.g. GRS1915+105 \citep{1999ApJ...513L..37M}, XTE J1550-564 \citep{1999ApJ...512L..43C}, GRO J1655-40 \citep{1999ApJ...522..397R}, or GX 339-4 \citep{2012A&A...542A..56N}),  or were estimated from the values holding for Sgr A* \citep{2006MNRAS.370..219M,2012NewA...17..254D}.
However, such parameters led to an extended accretion structure, meaning especially that $r_c^{\rm out} \sim 10^4M$ or more.

Here, we consider different values of parameters, and we perform higher dimensional simulations. 
We require that the outer critical point is located inside the computational domain in order to retain the flow subsonic at the outer boundary.
However, the total number of grid cells together with the sufficient resolution near the horizon restrict the maximal radius of the outer boundary, even though we use the logarithmic grid in radial direction.
Because of a sufficiently large value of the energy in the flow, $\mathcal{E}$, the radius $r_c^{\rm out}$ is located quite close to the centre.
The typical values of parameters used in this study are $\mathcal{E} = 0.0005, \gamma = 4/3, \lambda = 3.6M$. The shape of the corresponding solution is given in Fig. \ref{1D_sol}.

The evolution of initially non-magnetized gas is simulated with the \texttt{HARMPI} package supplied with our own modifications. The code conserves the vanishing magnetic field and there is no spurious magnetic field generated during the evolution.
We evolve two different types of initial conditions.
In the first case we prescribe $\rho$, $\epsilon$ and $u^r_{\tt BL}$ (radial component of the four-velocity) according to the Bondi solution, and we modify this solution by adding non-zero $u^\phi_{\tt BL}$ component of the four-velocity, where the $u^\alpha_{\tt BL}$ is the four-velocity in the Boyer-Lindquist coordinates.

The second option is to prescribe $\rho$, $\epsilon$ and $u^\alpha_{\tt BL}$ in accordance to the 1D shock solution (computed with Paczynski-Wiita potential in section \ref{s:Shocks}), and then follow the evolution.
Such solution obtained with the simplifying assumptions is of course not the true stationary solution. 
However, because of the fact, that for the given initial conditions and for chosen global parameters of the system, the gas evolves towards the true solution,
the discrepancies between the GR and PN are not expected to be substantial. Within the range of parameter space which
allows for a shock solution, this 1D approximation can be used for prescribing the initial conditions. 
We are interested in the evolution of such initial state towards  the correct stationary state.

The EOS of the form $p = (\gamma - 1) \rho \epsilon$ is used to close the system of equations, so that the isentropy assumption is not imposed. Hence, the specific entropy $K=p/\rho^\gamma$ is not constant and can evolve during the simulation.
The fiducial value of the polytropic index used in our computations is $\gamma=4/3$.

\subsection{Initial conditions -- Bondi solution equipped with angular momentum} \label{Ini_Bondi}

\begin{figure}
\begin{center}
 \includegraphics[width=0.5\textwidth]{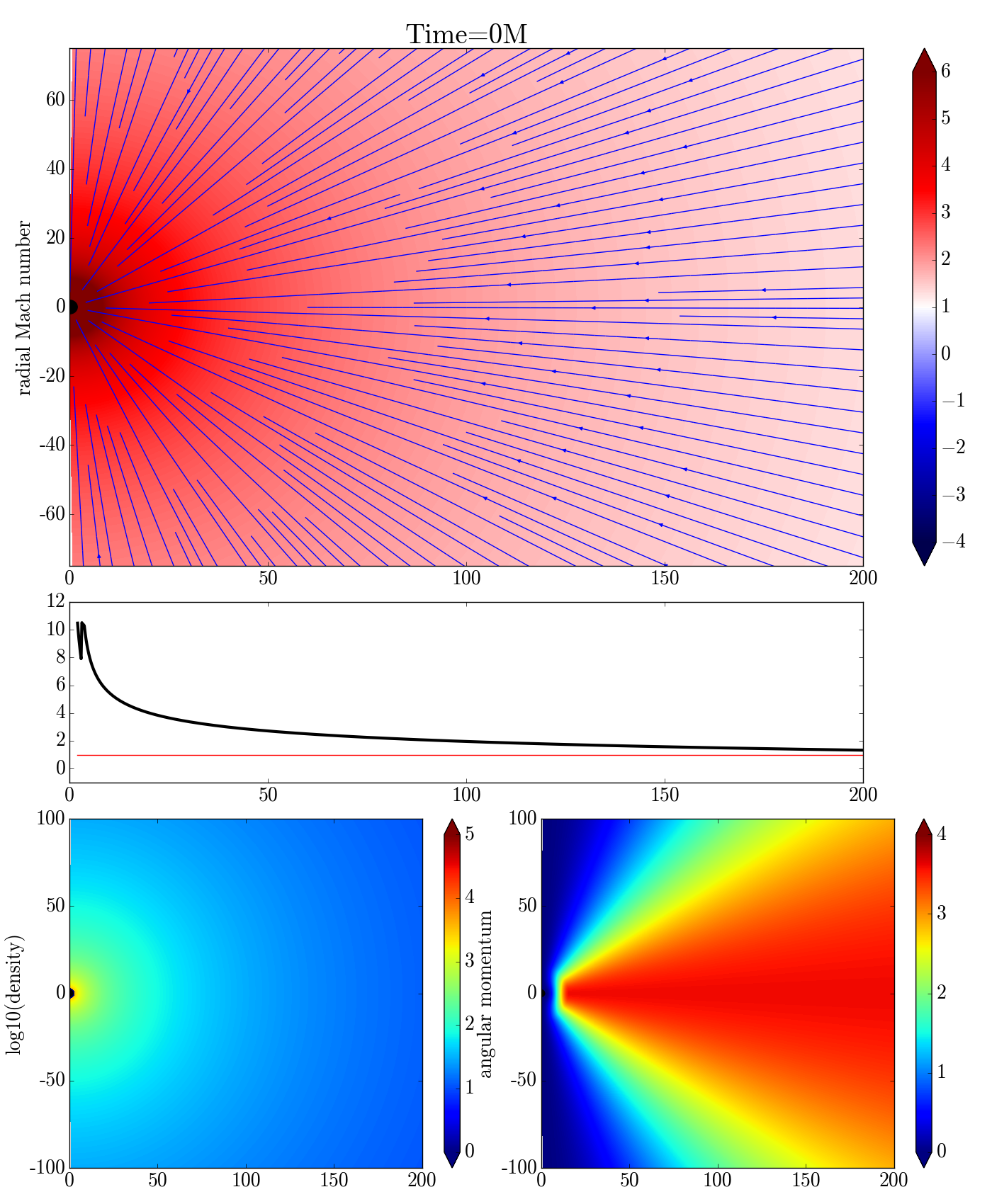}
 \caption{Model \mD{}: Bondi initial conditions with $\lambda^{\rm eq}=3.6$M.  The set of four panels shows the initial conditions at $t=0$, in particular the slices of $\mathfrak{M}$ with streamlines of the flow plotted (top panel), its equatorial profile (middle panel), and distributions of the flow density $\rho$ and angular momentum $\lambda$ (bottom panel).
\label{BondiRotInit} }
 \end{center}
\end{figure}

In our first set of computations we adopted initially $\rho$, $\epsilon$ and $u^r_{\tt BL} = -v$  according to the Bondi solution with $ \epsilon = K \rho^{\gamma-1} /(\gamma-1)$, where $K$ is given by Eq. (\ref{konst_K}) evaluated for the Bondi critical point (solution of Eq. (\ref{r_c}) with $\lambda = 0$). The solution is parametrized by the value of the polytropic exponent~$\gamma$ and the energy $\mathcal{E}$, which fix the position of the critical point $r_c^{\rm out}$.
We modify the initial conditions by prescribing the rotation according to relations
\begin{eqnarray}
\lambda = \lambda^{\rm eq} \sin^2{\theta}, \qquad r&>&r_{\rm b}, \label{uphi}\\
\lambda = 0 ,\qquad r&<&r_{\rm a}, \label{uphi_null}
\end{eqnarray}
in the Boyer-Lindquist coordinates. Between $r_{\rm a}$ and $r_{\rm b}$ the values are smoothened by a cubic spline. The time component of four-velocity is set from the normalization condition  $g_{\mu\nu}u^\mu u^\nu = -1$ assuming $u^\theta_{\tt BL}=0$.
The factor $\sin^2 \theta$ in Eq. (\ref{uphi}) ensures that angular momentum vanishes smoothly at the axis, hence the maximal value of angular momentum $\lambda^{\rm eq}$ is  achieved only in the equatorial plane.
One example of such initial conditions is plotted in Fig.~\ref{BondiRotInit}.

\subsection{Initial conditions -- Shock solution} \label{Ini_shock}
In the second type of simulations we modify the initial data procedure such that we find the solution with the shock in the same way as in \cite{our_paper}.
The values of $\rho$, $\epsilon$ and $u^r_{\tt BL}$ are set accordingly, with $ \epsilon = K^{\rm in/out} \rho^{\gamma-1} /(\gamma-1)$, where $K^{\rm in/out}$ is given by Eq. (\ref{konst_K}) evaluated at the corresponding critical point $r_c^{\rm in/out}$ for the two branches of solution.

\begin{figure}
\begin{center}
 \includegraphics[width=0.5\textwidth]{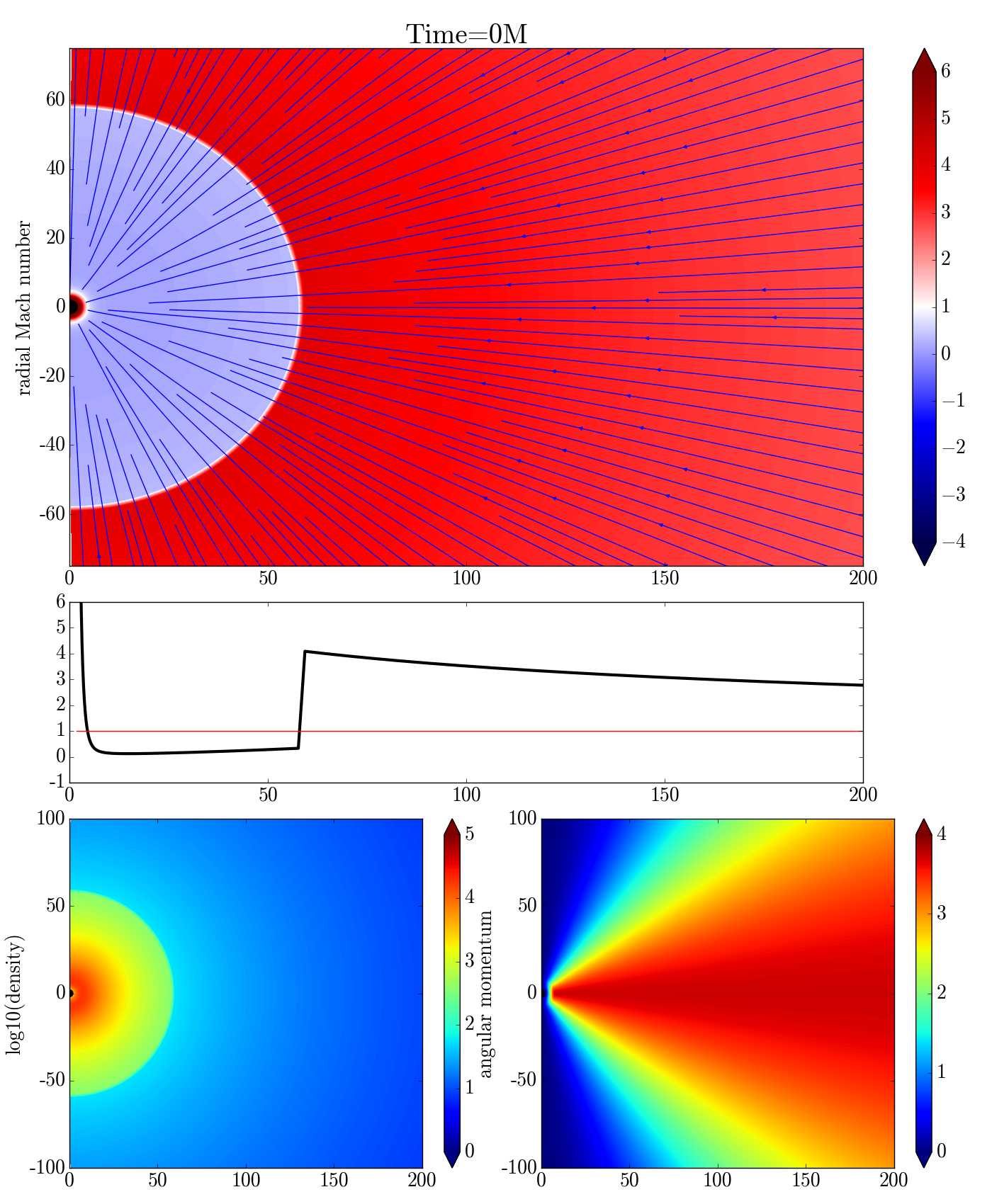}
  \caption{ Model \mI{}: Initial data with a shock with $\mathcal{E}=0.0005, \lambda^{\rm eq}=3.72$M. \label{K112_Mach_Ini}}
 \end{center}
\end{figure}

However, the 1D analysis provided in that paper and here in Section~\ref{s:Shocks} was based on pseudo-Newtonian approximation, while now we use GR MHD code in order to examine the differences between the two approaches. Moreover, the 1D analysis was held under the assumption of the quasi-spherical shape of the flow and the constant value of angular momentum.
That in particular means, that the dependence of the mass accretion rate, which is constant along the flow, scales with $r^2$.
We cannot straightforwardly extend  this model into 3D, in other words we can not set the spherical distribution of matter with the angular momentum which would be constant everywhere, because we need to avoid a non-zero angular momentum near the vertical axis.
However, the choice of the angular momentum profile in $\theta$ direction is arbitrary, if it drops to zero near the axis.
Therefore, we choose two different profiles of angular momentum for the initial and boundary conditions, to see how much the results depend on this distribution.

\subsubsection{Angular momentum scaled by $\sin^2 \theta$} \label{Ini_shock_sph}
The first choice is the same as in Section~\ref{Ini_Bondi}, given by Equations (\ref{uphi}) and (\ref{uphi_null}).
In this case, the maximal value of angular momentum $\lambda^{\rm eq}$ is obtained only in the equatorial plane and it is lower elsewhere.
This type of initial conditions is shown on Fig.~\ref{K112_Mach_Ini}.

\subsubsection{Constant angular momentum in a cone} \label{Ini_shock_cone}

\begin{figure}	
 \includegraphics[width=0.5\textwidth]{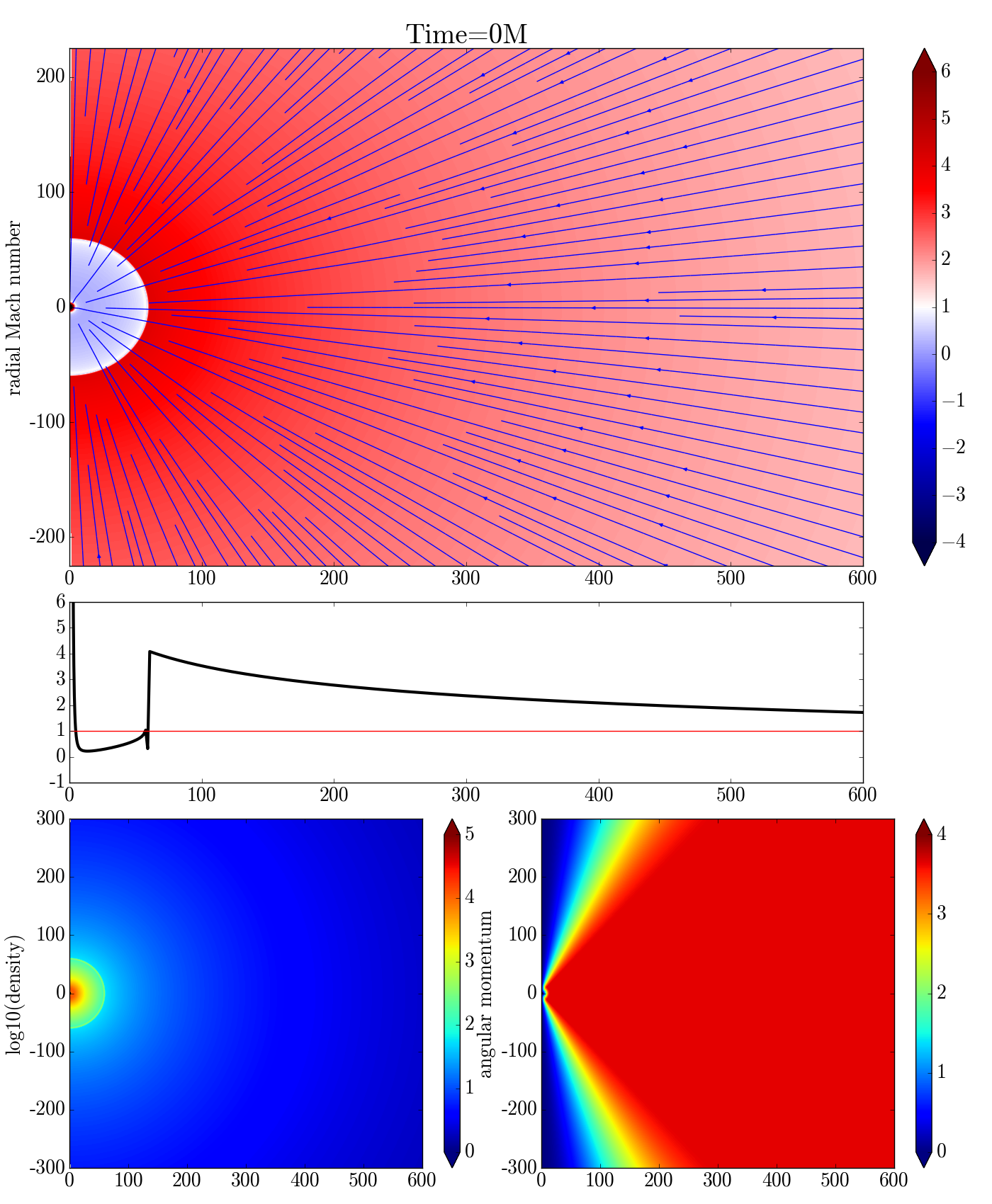}
   \caption{ Initial data for the shock and constant angular momentum in a cone, $\mathcal{E}=0.0005, \lambda^{\rm eq}=3.65$M, $\theta_c=\pi/4$ (model \mHc).   \label{K430_Mach_Ini} }
\end{figure}
In the second case we prescribe constant angular momentum in a cone with the half-angle $\theta_c$ centered along the equatorial plane. 
The values of $\lambda$ are smoothed down from the cone towards the axis. For this smoothening we choose a  cubic spline  given by the relations:
\begin{eqnarray}
\lambda = \lambda^{\rm eq} f \qquad \qquad \quad \qquad && \\
f = \frac{\theta^2(3\theta_c - 2\theta)}{\theta_c^3}, \quad \theta<\frac{\pi}{2}-\theta_c \\
f = 1, \quad \theta\in[\frac{\pi}{2}-\theta_c,\frac{\pi}{2}+\theta_c] \label{lambda-cone}\\
f= \frac{(\theta-\pi)^2(\frac{\pi}{2} + 3\theta_c - 2\theta)}{(\theta_c-\frac{\pi}{2})^3}, \quad \theta>\frac{\pi}{2}+\theta_c
\end{eqnarray}

Thus, all gas within the cone has the maximum angular momentum of $\lambda^{\rm eq}$, which resembles more the assumptions of the 1D model.
We show the initial conditions for this case in Fig.~\ref{K430_Mach_Ini}.

\ \newline

Because of these modifications of angular momentum distribution, such initial conditions are not expected to be the stationary solution in higher dimensions.
However, our experience with 1D simulations shows, that only the presence of the inner sonic point is essential for the creation of the shock in the flow, and the exact stationary solution is not needed.
If the inner sonic point is present, then the shock bubble shape adjusts itself after a short transient time into the appropriate form.

\begin{figure}
 \includegraphics[width=0.48\textwidth]{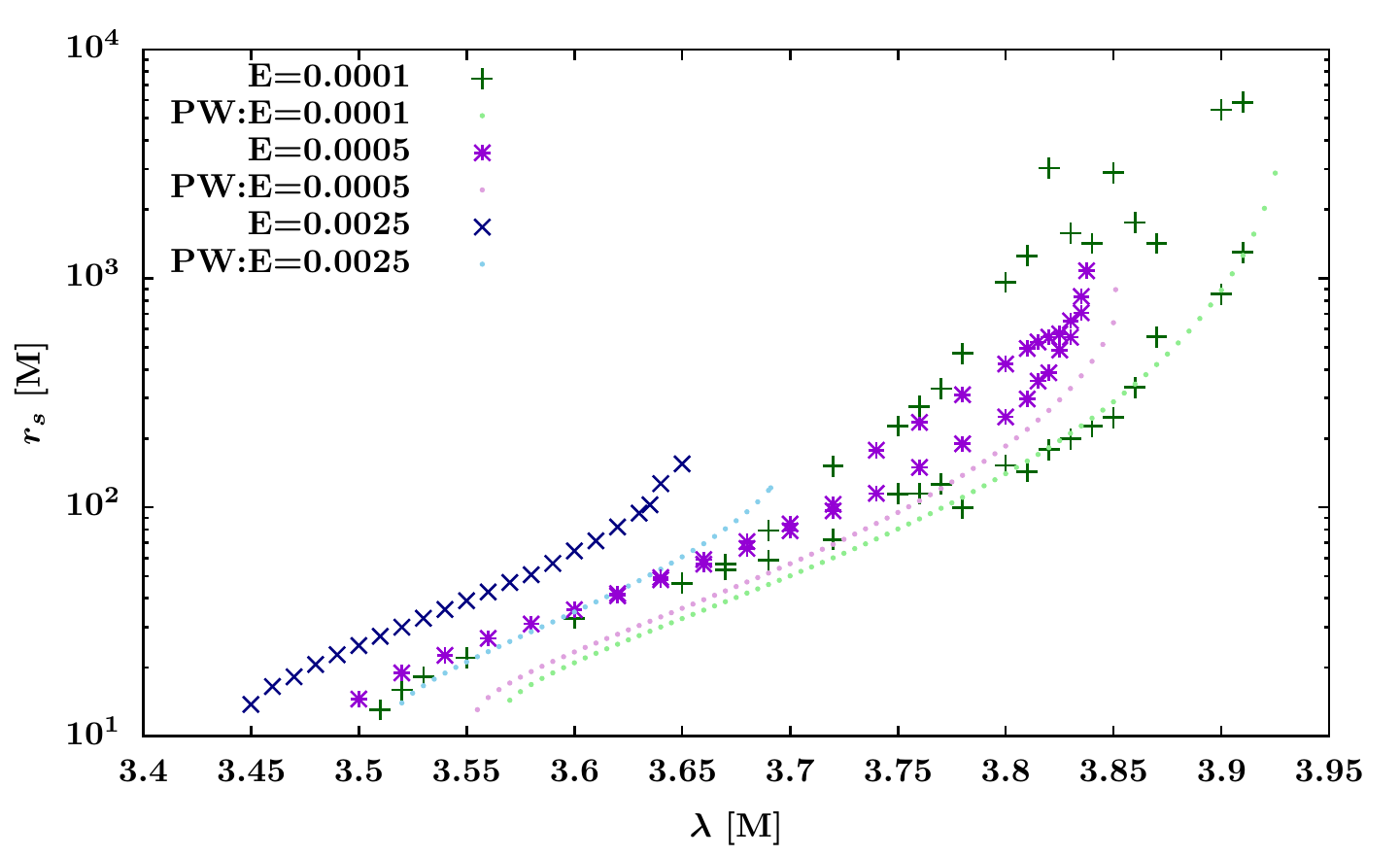}
  \includegraphics[width=0.48\textwidth]{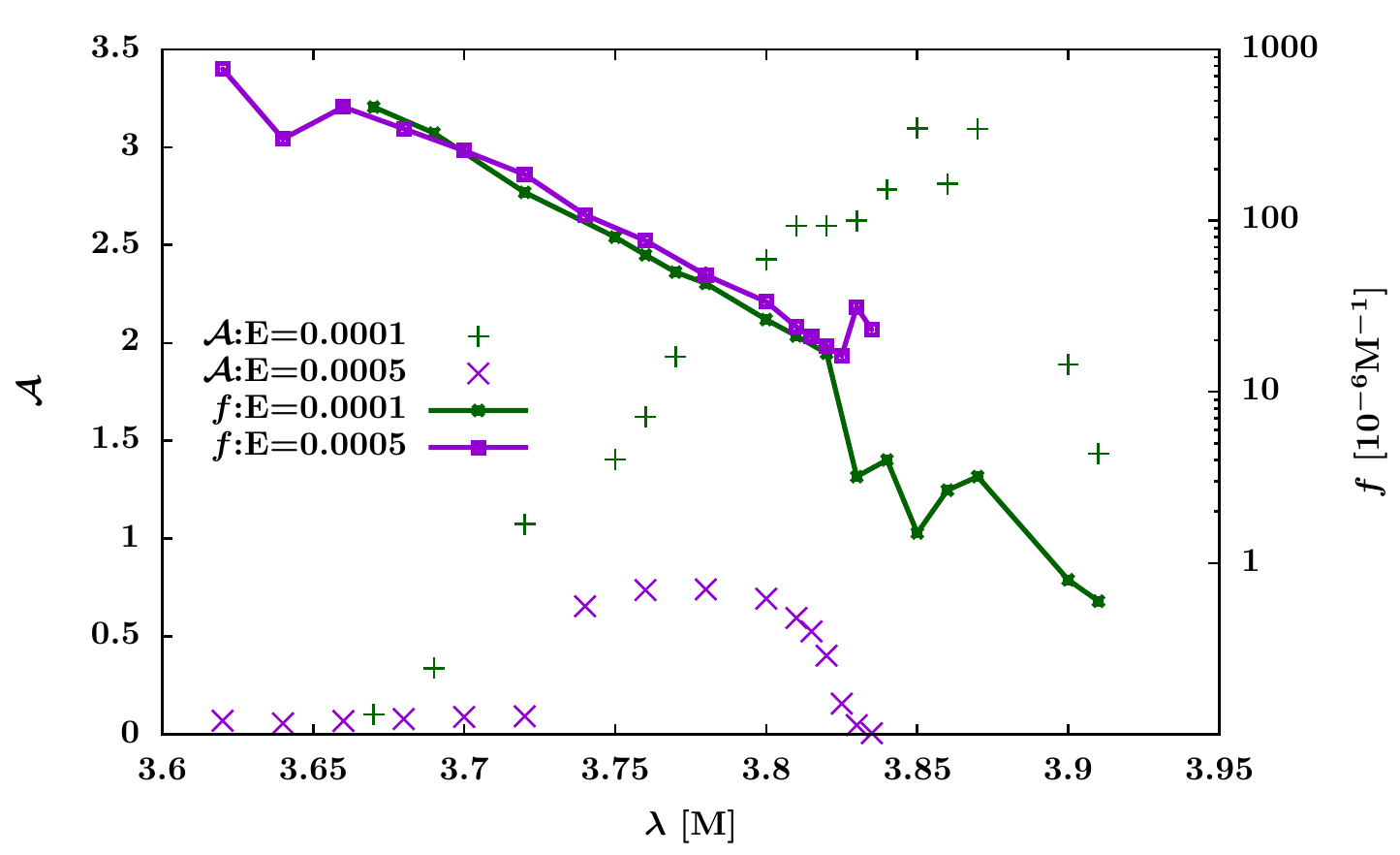}
   \caption{ Top panel: Position of the shock front depending on angular momentum for different values of energy. In case of oscillations, the minimal and maximal shock position is shown. Comparison with the shock front position obtained in the semi-analytical approach using Paczynski-Wiita potential is shown with small dots.  Bottom panel: The amplitude $\mathcal{A}$ of the oscillations of the mass accretion rate and its frequency  for the oscillating models. \label{1D_shock} }
\end{figure}

\subsection{Initial conditions for spinning black hole}\label{Ini_spin}
For a spinning black hole we use the initial data described in \ref{Ini_shock_sph} (angular momentum is scaled by $\sin^2\theta$).

Our semi-analytical 1D shock solution is obtained for the Schwarzschild black hole only. In the case of spinning black hole, the relevant values of angular momentum for the shocks can vary significantly and can be outside the possible existence of shocks in the Schwarzschild space time.
Hence, we use two different values of angular momentum: (i) $\lambda^s$, to find the semi-analytic shock solution in non-spinning space time, according to which $\rho$, $\epsilon$, $u^r_{\tt BL}$ and $K$ are set, and (ii) $\lambda^{\rm eq}_g$, which is a different value, according to which the rotation is prescribed using Equations (\ref{uphi}) and (\ref{uphi_null}), and the normalization condition for the four-velocity.

The value of $\lambda^s$ is not important for the long term evolution; it only enables us to find a configuration with shock which is used for the initial conditions, so that there exists an inner sonic point at the initial time.
However, the gas is rotating with  $\lambda^{\rm eq}_g$, which determines the evolution of the flow. After a short transition time, the other variables distribution adjusts to the actual angular momentum.

\section{Results} \label{s:Results}

\subsection{1D computations}\label{results_1D}
In the general relativistic framework, the local sound speed relative to the fluid is given by
$$
c_s ^{2} = \gamma p \left(\rho +\frac{ \gamma p}{\gamma -1}\right)^{-1},
$$ and the radial Mach number is obtained as $\mathfrak{M} = - u_{\rm BL}^{\tt r}/c_{\rm s}$.
The sonic points are the points, where the flow smoothly passes from subsonic to supersonic motion, that is $\mathfrak{M} = 1$, and its value increases inwards.
The shock front is at the place, where $\mathfrak{M}$ discontinuously changes from $\mathfrak{M}>1$ to $\mathfrak{M}<1$ along the flow (with decreasing $r$).

\subsubsection{Properties of the flow for a constant angular momentum}

\begin{figure}
 \includegraphics[width=0.48\textwidth]{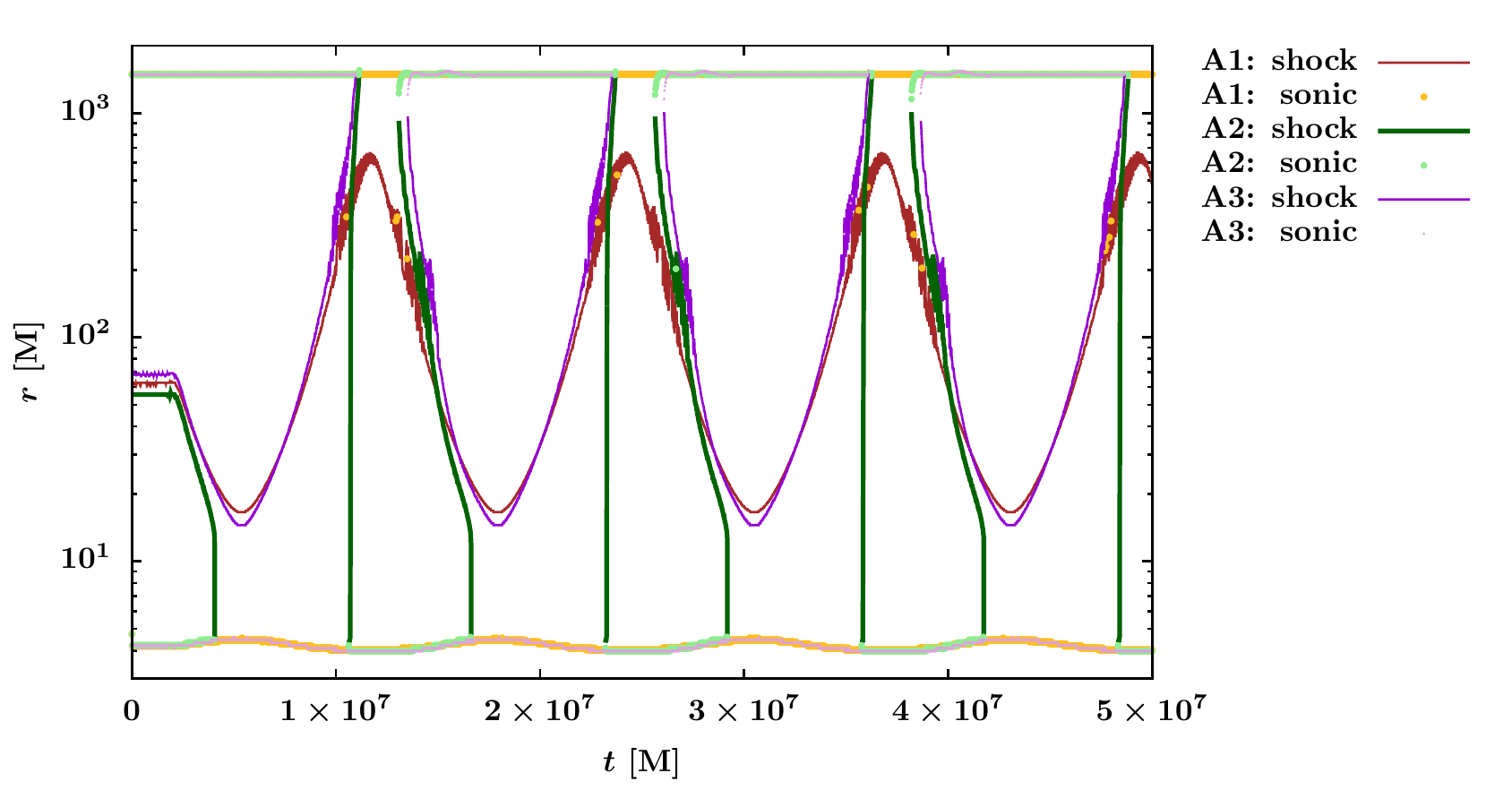}
   \caption{ Position of the shock front and the sonic points for three different models with changing angular momentum at the outer boundary as a function of time. Model \mAa{}: $\lambda^{\rm eq} (0)=3.67$M, $A=0.16$M, $P=2\cdot10^6$M, model \mAb{}: $\lambda^{\rm eq} (0)=3.655$M, $A=0.2$M, $P=2\cdot10^6$M, model \mAc{}: $\lambda^{\rm eq} (0)=3.68$M, $A=0.18$M, $P=2\cdot10^6$M. All models have $\mathcal{E}=0.005, \gamma=4/3$.\label{1D_loop} }
\end{figure}

The resulting profile of the Mach number $\mathfrak{M}$ for the converged stationary state is shown in Fig.~\ref{1D_sol} for four different values of $\mathcal{E}$. The inferred shock and outer sonic point locations are given in Table~\ref{t:1D-shock}.

\begin{table}
\begin{tabular}{c|cccc}
$\mathcal{E}$ & 0.00002 & 0.0001 & 0.0005 & 0.0025 \\ \hline
$r_s$ & 31 & 32 & 36 & 64 \\
$r_{\rm son}^{\rm out}$ & $3.7 \cdot 10^4$ &  $7.5 \cdot 10^3$ & $1.5 \cdot 10^3$ & $2.8 \cdot 10^2$\\
$\mathcal{R}$&10&9.4&7.6&4\\
\end{tabular}
\caption{Location of shock front and the outer sonic point of stationary solutions for different values of $\mathcal{E}$. The values in geometrized units ([M]) are inferred from the solution at the end of the simulation, at $t_f=10^6$M.  \label{t:1D-shock}}
\end{table}

For higher energy, the outer sonic point is closer to the black hole unlike the shock position, which is located farther, and the outer supersonic region of the flow is thus shrinking.
The strength of the shock is anti-correlated with the energy (and with the location of the shock), so that for increasing energy the ratio of the post-shock density to the preshock density ($\mathcal{R}$) is decreasing (see Table~\ref{t:1D-shock} and panel b) in Figure~\ref{1D_sol}).

In comparison with the pseudo-Newtonian analytical estimate, which put the shock position for $\mathcal{E}=0.0001$ at $r_s^{\rm PW}(0.0001) = 21M$, the GR computation tends to put the shock farther from the black hole.
On the other hand, the minimal stable shock position is very similar, because in GR computations the shock exists for slightly lower values of $\lambda$, which can be seen on Fig.~\ref{1D_shock}. 
Here the dependence of the shock front position on angular momentum is shown for both the pseudo-Newtonian and GR computations.
The radial extend of possible shock existence agrees very well between PW and GR results, however for the same value of angular momentum the GR shock is located farther from the black hole.

Similarly like in the case of PW simulations, which we presented in \cite{our_paper}, also in the GR case computed with \texttt{HARMPI} we have found oscillations of the shock front for higher angular momentum, which causes also the oscillation of the mass accretion rate through the inner boundary. In Fig.~\ref{1D_shock} in the top panel we show the minimal and maximal shock positions during the simulation for the oscillating cases. In the bottom panel, we show the amplitude of the oscillations of the mass accretion rate, which is computed as the ratio of the difference between the maximal and minimal accretion rate to its mean value $\mathcal{A} = ({\rm max}(\dot{M}) - {\rm min}(\dot{M}))/\bar{\dot{M}}$. This ampplitude and the corresponding frequency is presented for the oscillating models.

\begin{table}
\begin{tabular}{c|c|ccccccc}
\multirow{2}{*}{$\mathcal{E}_1$}&$\lambda$ & 3.20 & 3.21 & 3.22 & 3.25 & 3.30 & 3.42 & 3.43 \\
&$r_s$ & -- & 11.9 & 14.0 & 19.8 & 31.6 & 172.9 & -- \\ \hline

\multirow{2}{*}{$\mathcal{E}_2$}&$\lambda$ & 3.245 & 3.25 & 3.26 & 3.30 & \multicolumn{2}{l}{3.40}  \\
&$r_s$ & -- & 10.7 & 13.3 & 20.7 & \multicolumn{2}{l}{46 -- 99}
\end{tabular}
\caption{ Location of shock front, if existent, with $a=0.3,\gamma=4/3$ for different values of $\lambda$ for $\mathcal{E}_1=0.002$ and $\mathcal{E}_2=0.0000033$. The values in geometrized units ([M]) are inferred from the solution at the end of the simulation. \label{t:1D-Das}}
\end{table}

The general relativistic semi-analytical study of the shock solutions in Kerr metric was given in \cite{2012NewA...17..254D}, however those authors considered the disc in hydrostatic equilibrium with vertically averaged quantities.
The shape of the different regions in the parameter space is given in  \cite{2012NewA...17..254D}, Fig.~1 for $a=0.3$.
On Fig.~3 in that paper, the authors show the parameter space of possible shock existence for $\tilde{\mathcal{E}}=1.0000033, \gamma=4/3, a=0.3$, where $\tilde{\mathcal{E}}$ is the total specific energy and corresponds to $\tilde{\mathcal{E}} = \mathcal{E}+1$.
To compare their solutions with our results, we performed a set of simulations with $a=0.3$ for $\mathcal{E}_1=0.002$ and $\mathcal{E}_2=0.0000033$. The results are summarised in Table~\ref{t:1D-Das}.

 For $\mathcal{E}_1$   \cite{2012NewA...17..254D} predicts that the shock exists in the subset of the region A (in particular in the region A$_{\rm S}$, which is not shown in the figure), which gives approximately the range 2.8 M $< \lambda <$ 3.08 M.
Our simulations show the shock existence for higher values of angular momentum, in particular the shock front is accreted up to $\lambda=3.2$ M and the stationary shock exists for $\lambda \in [3.21 \rm{M},3.42 \rm{M}]$.

For $\mathcal{E}_2$, those authors predict the shock existence for $\lambda > 3.239$M.
We have found the stable shock solution for $\lambda=3.25$~M, while for $\lambda=3.245$~M the shock is accreted.
When the angular momentum is increased, the oscillations develop, for $\lambda=3.4$~M their frequency $f\sim 2.9\cdot 10^{-4}$~M$^{-1}$ with the amplitude $\mathcal{A} = 1.45$.

Because a different configuration is considered in the two papers (disc in vertical hydrostatic equilibrium versus quasi-spherical flow),
some differences are expected and they are more prominent for higher energy of the flow.

Quite recently, \cite{2016arXiv161206882T} studied the low angular momentum flows with shocks in the Kerr spacetime in several different geometries with different physical conditions at the shock front. They included also the case of the conical flow with energy-preserving shock, which is close to our scenario. Figure 2 of \cite{2016arXiv161206882T} shows, that indeed the multicritical region for the quasi-spherical flow exists for higher values of angular momentum than for the flow in vertical hydrostatic equilibrium. The quantitative comparison of our results with this paper will be given in the future work.

\subsubsection{Properties of the flow for a time dependent angular momentum}

\begin{figure}
\begin{center}
 \includegraphics[width=0.5\textwidth]{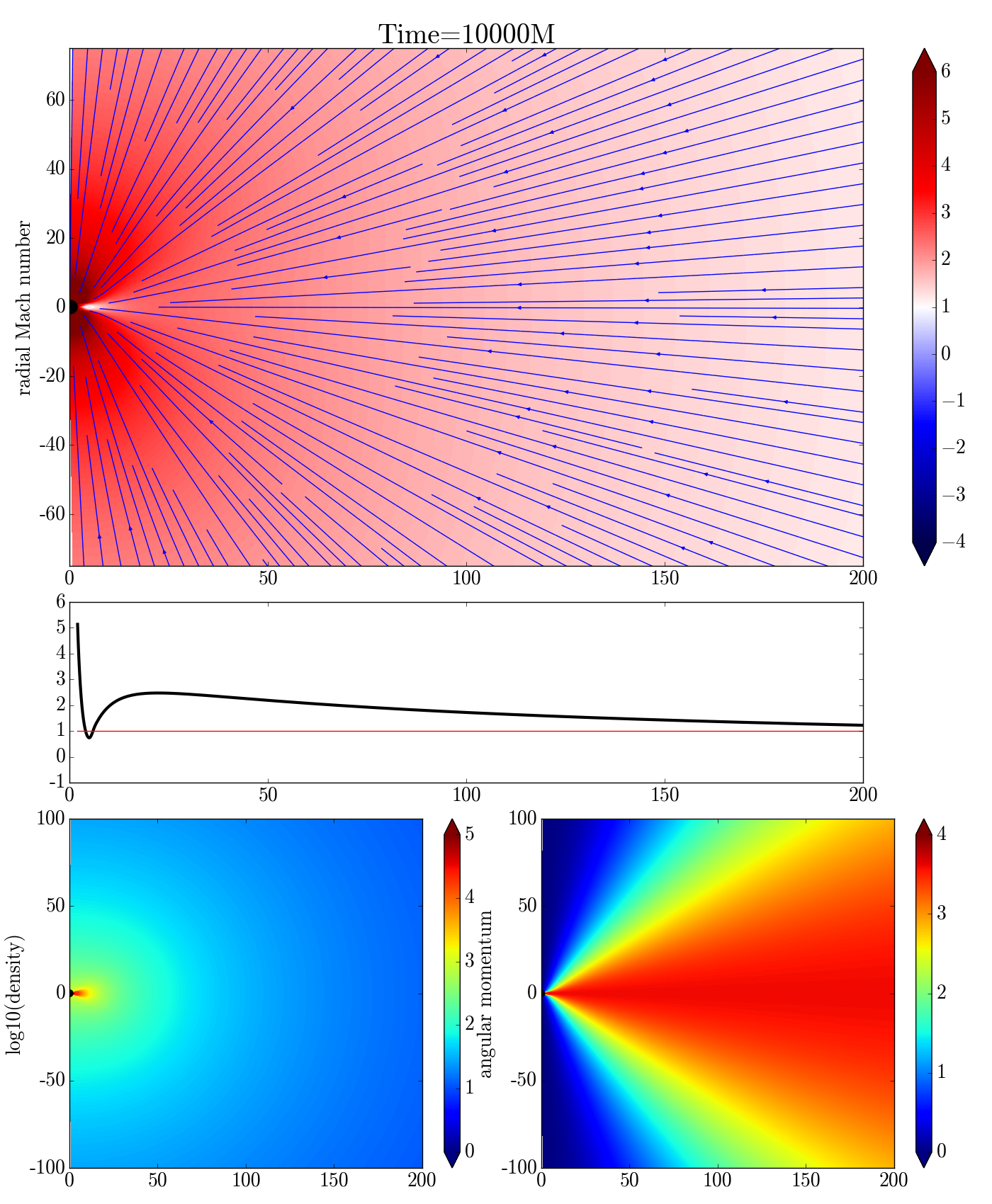}
 \caption{Model \mD{}: The snapshot at the end of the simulation shows the evolved gas at $t=10^4$M, which settles into the outer branch, no shock appears during the evolution.   \label{BondiRot} }
 \end{center}
\end{figure}

Further, we repeated the PW computations done previously with the code \texttt{ZEUS} \citep{our_paper},
regarding the hysteresis loop connected with the shock existence with changing angular momentum of the incoming matter.
In this case we prescribe the angular momentum of the matter coming through the outer boundary according to the time dependent relation:

\begin{equation}
\lambda^{\rm eq} (t) = \lambda^{\rm eq} (0)  - A\sin(t/P),
\end{equation}
where $A$ is the amplitude and $P$ is the period of the perturbation. If we choose $A$ and $P$ such that the angular momentum crosses the boundary of the multicritical region from below and from above, we observe the creation and disappearance of the shock front, similarly like it was in the PW case.

The comparison of the shock front movement for three different models is given in Figure~\ref{1D_loop}. Model \mAa{} with  $\lambda^{\rm eq} (0)=3.67$~M, $A=0.16$~M, $P=2\cdot10^6$~M does not exceed the boundary of the shock existence interval from neither side. The shock is moving in accordance with the changing angular momentum to and from the black hole, spanning the region (16.6~M, 672.5~M). Both sonic points exist persistently during the evolution.

Model \mAb{} with $\lambda^{\rm eq} (0)=3.655$M, $A=0.2$M, $P=2\cdot10^6$M crosses the shock existence boundary on both sides.
As a consequence, the shock front is being accreted very quickly after it reaches the minimal stable shock position.
The time scale of this event is given by the advection time from the minimal stable shock position and does not depend on the perturbation parameters $A$ and $P$.

From that moment the inner sonic point vanishes and the flow follows the outer Bondi-like branch of solution (for which only the outer sonic point exists) until the time, when the angular momentum increases such that the outer solution no longer exists.
At that point, the shock is formed at the position of the inner sonic point and it is moving very fast outward, where it merges with the outer sonic point, so that the type of accretion flow with the inner sonic point only is established. Later, when the angular momentum of the flow is decreasing again, the shock forms at the outer sonic point and it moves slowly towards the black hole.

The third example (model \mAc{} with $\lambda^{\rm eq} (0)=3.68$M, $A=0.18$M, $P=2\cdot10^6$M) shows the case, where only the upper boundary is crossed.
Here we have the shock moving slowly towards and away from the centre, where it merges with the outer sonic point. For some time only the accretion with the inner sonic point exists and then the shock appears again. In this case, the timescale of such changes and the velocity of the shock front is uniquely given by the parameters $A$ and $P$.

\begin{figure}
\begin{center}
 \includegraphics[width=0.5\textwidth]{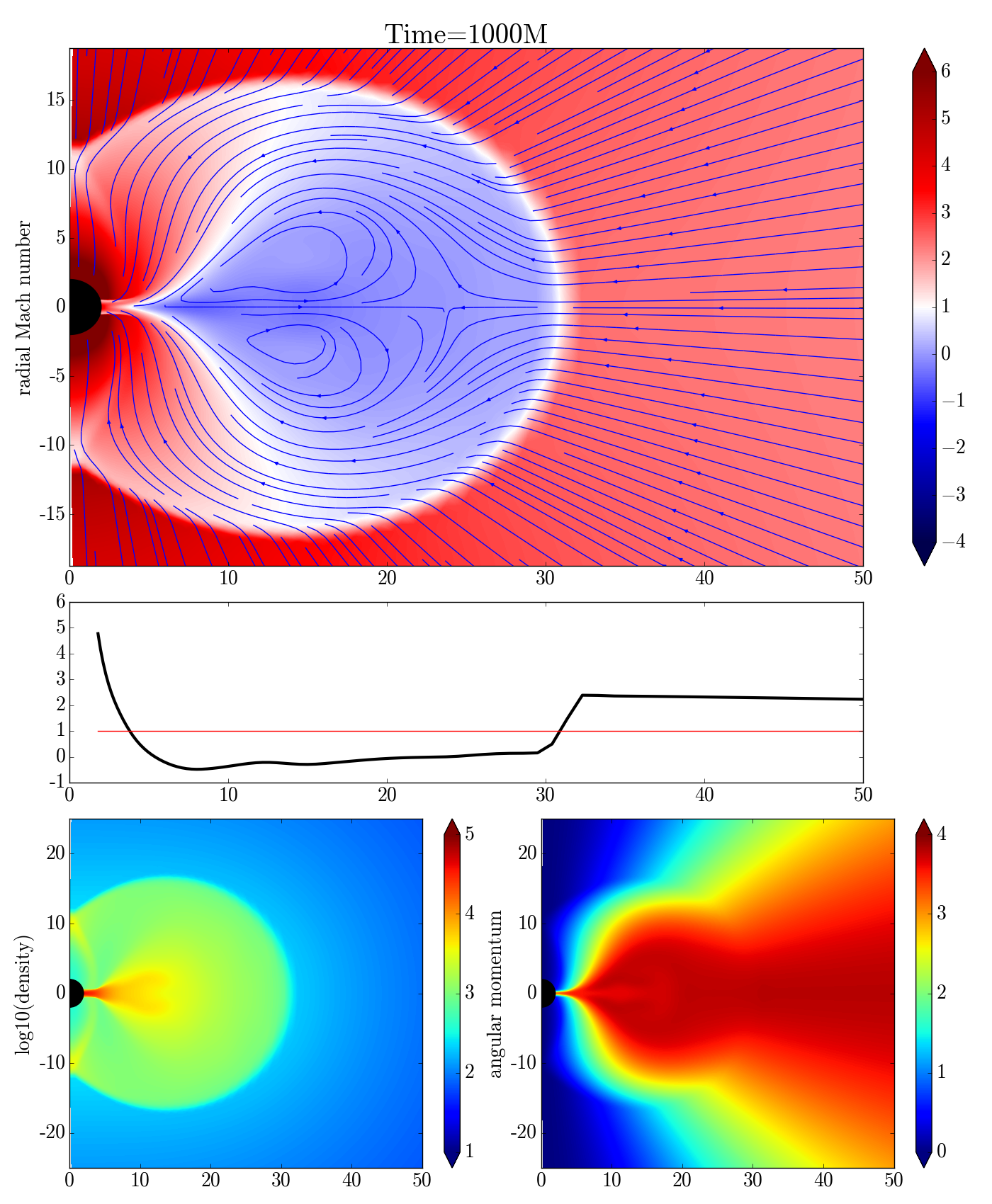}
 \includegraphics[width=0.5\textwidth]{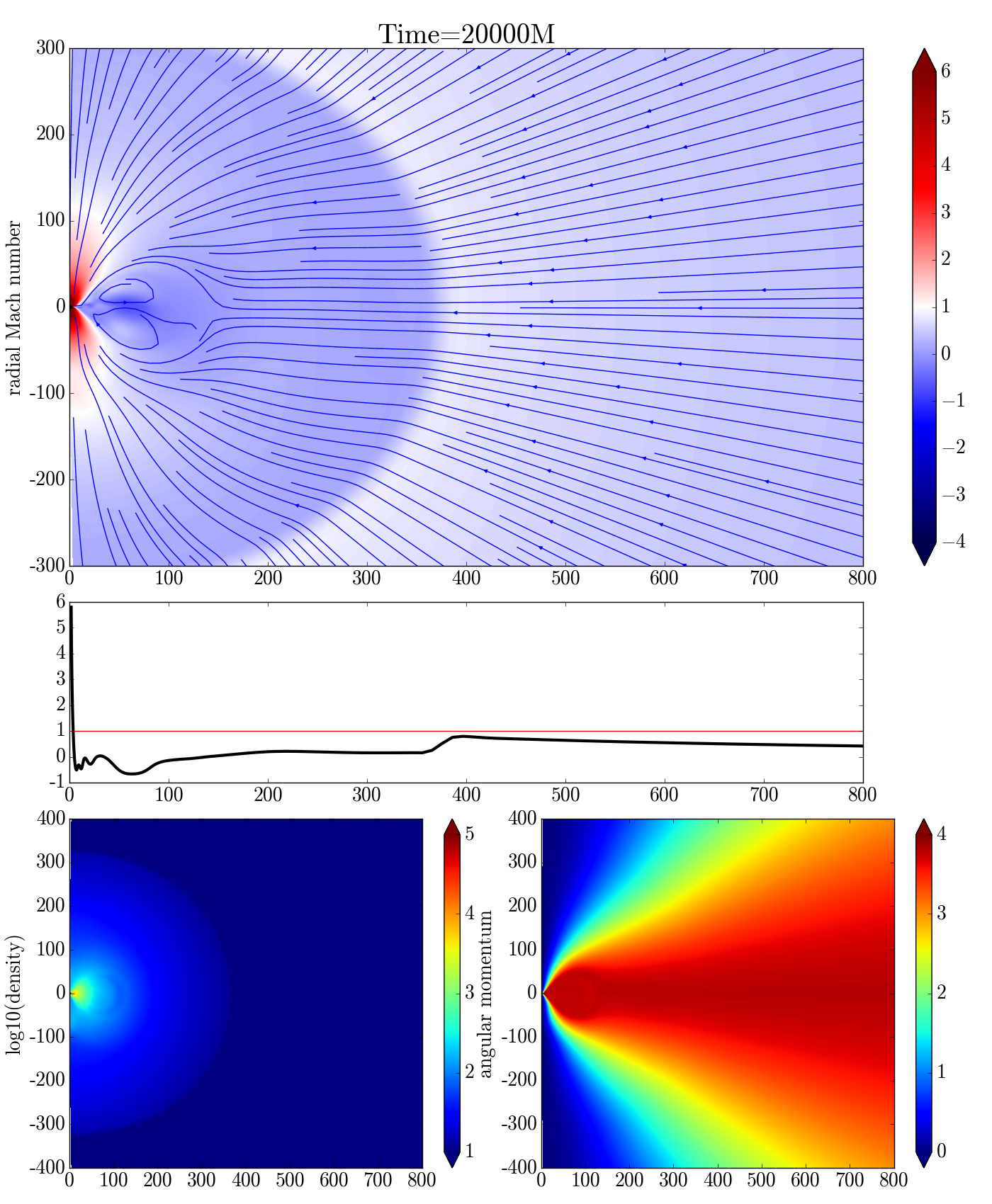}
 \caption{ Model \mE{}: Bondi initial data with $\mathcal{E}=0.0025, \lambda^{\rm eq}=3.8$M. The shock is expanding (the first snapshot at $t=10^3$M) until it merges with the outer sonic surface (second snapshot at $t=2 \cdot 10^4$M).  Note the different spatial range of the second snapshot. \label{B10_II} }
 \end{center}
\end{figure}

The case, in which only the lower boundary is crossed, is not shown, because in this case the shock front is accreted during the first cycle and never appears again.

Hence, under certain circumstances, the standing shock can appear in the low angular momentum flow during accretion. Then it can either (i) stay at certain position, (ii) move slowly downwards or upwards in the accretion flow with velocity given by the rate of change of angular momentum given by parameters $A$ and $P$ in our model, (iii) be accreted quickly from the the minimal stable shock position (close to the black hole), or (iv) be formed close to the black hole and move quickly towards the outer sonic point.
In the last two cases, the velocity of the shock front does not depend on the perturbation parameters $A$ and $P$, but it is given by the properties of the medium. We will study this issue in the next section with 2D simulations; in general, the velocity of the shock front is a few times lower than the sound speed in the preshock medium.

\begin{figure}
\begin{center}
 \includegraphics[width=0.5\textwidth]{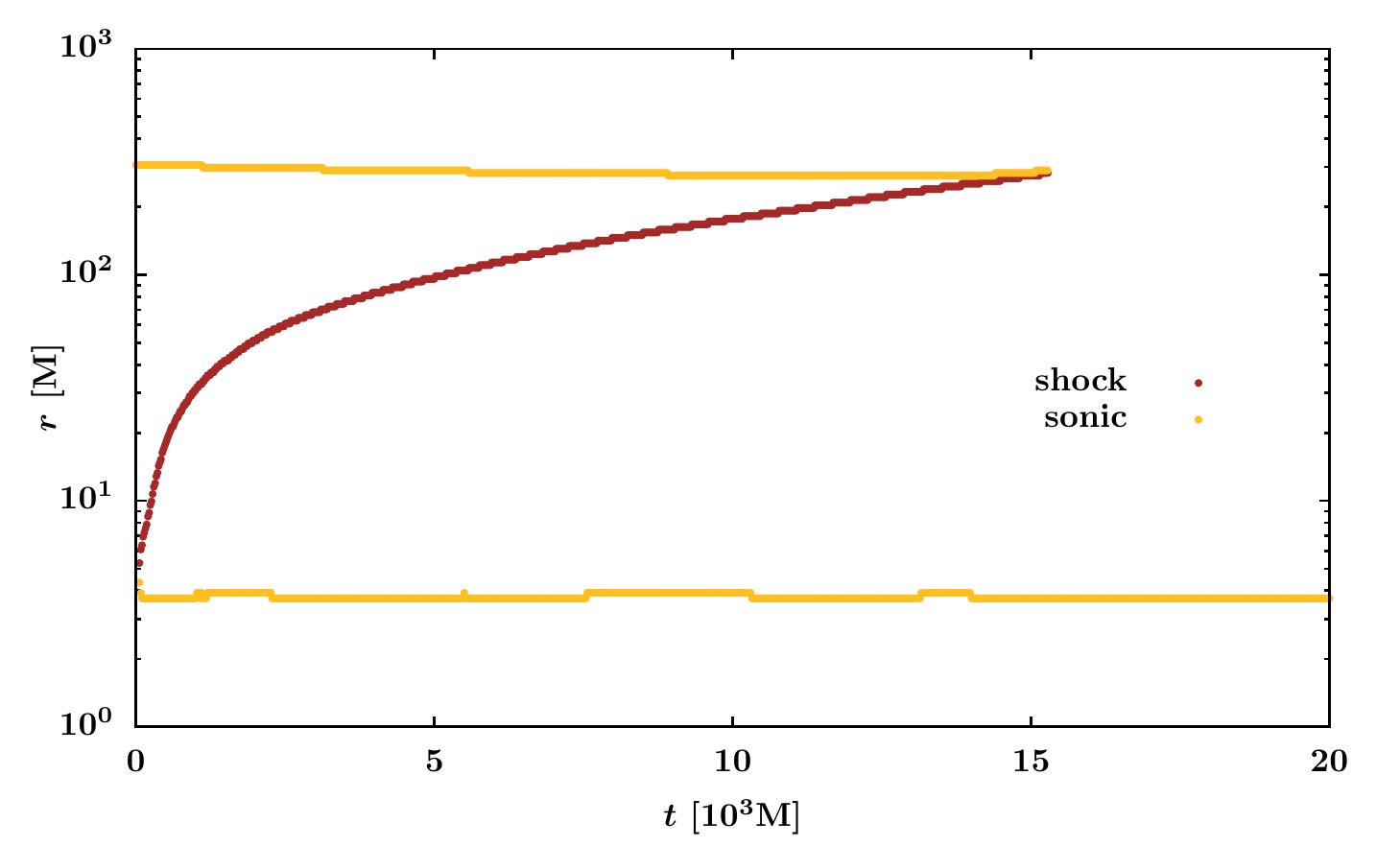}
 \caption{ Model \mE{}: expanding shock with Bondi initial data. We track the position of the transonic points (i.e. $(\mathfrak{M}[i] -1)(\mathfrak{M}[i+1] -1) <1$) in the equatorial plane ($i$ is the index of the radial coordinate on the grid) and we labeled the sonic points  ( $\mathfrak{M}[i]>\mathfrak{M}[i+1]$)
 by yellow points and the shock position ( $\mathfrak{M}[i]<\mathfrak{M}[i+1]$) by brown points.
  \label{B10_II_rs} }
 \end{center}
\end{figure}

\begin{figure}
 \includegraphics[width=0.48\textwidth]{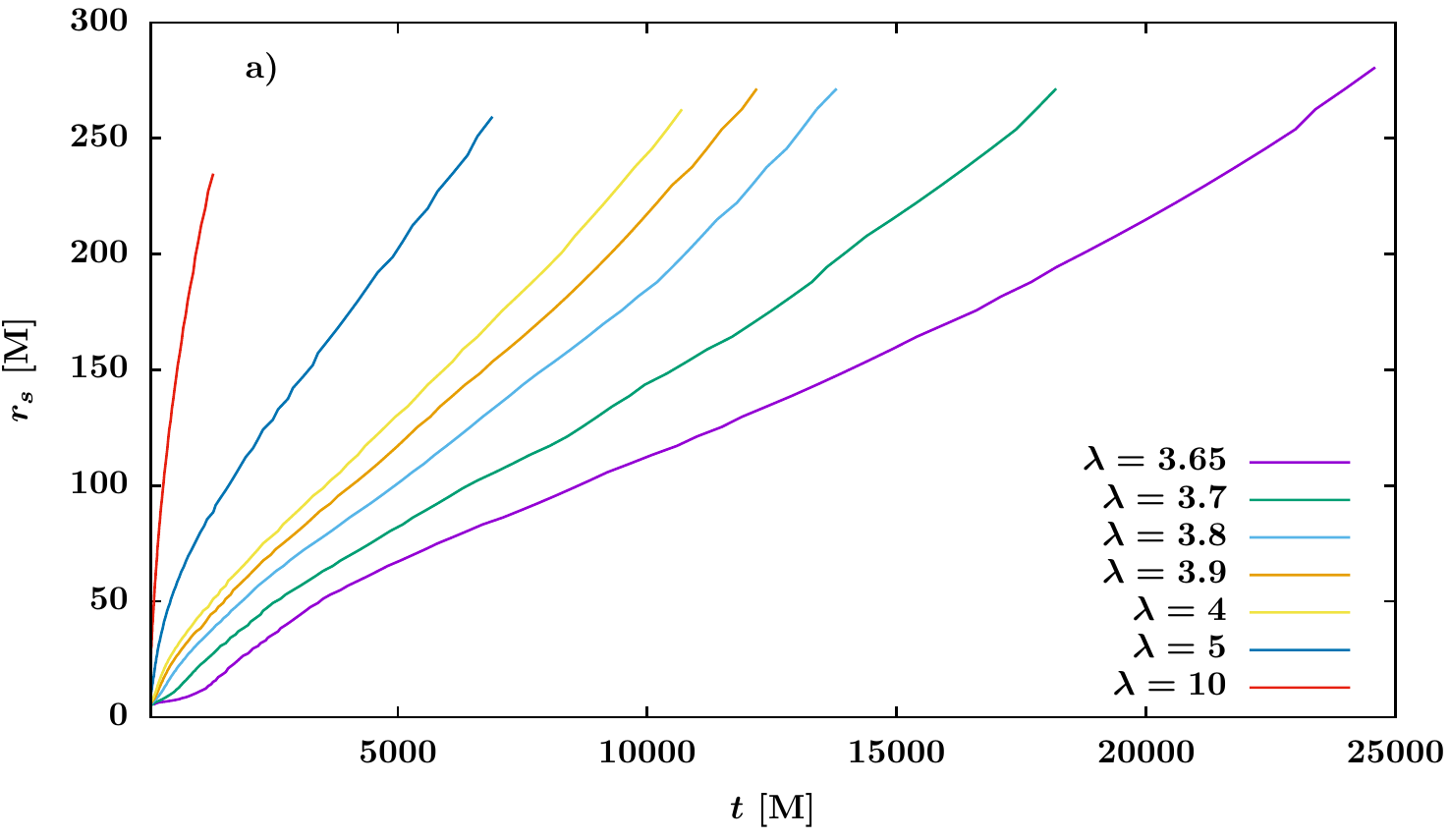}
 \includegraphics[width=0.48\textwidth]{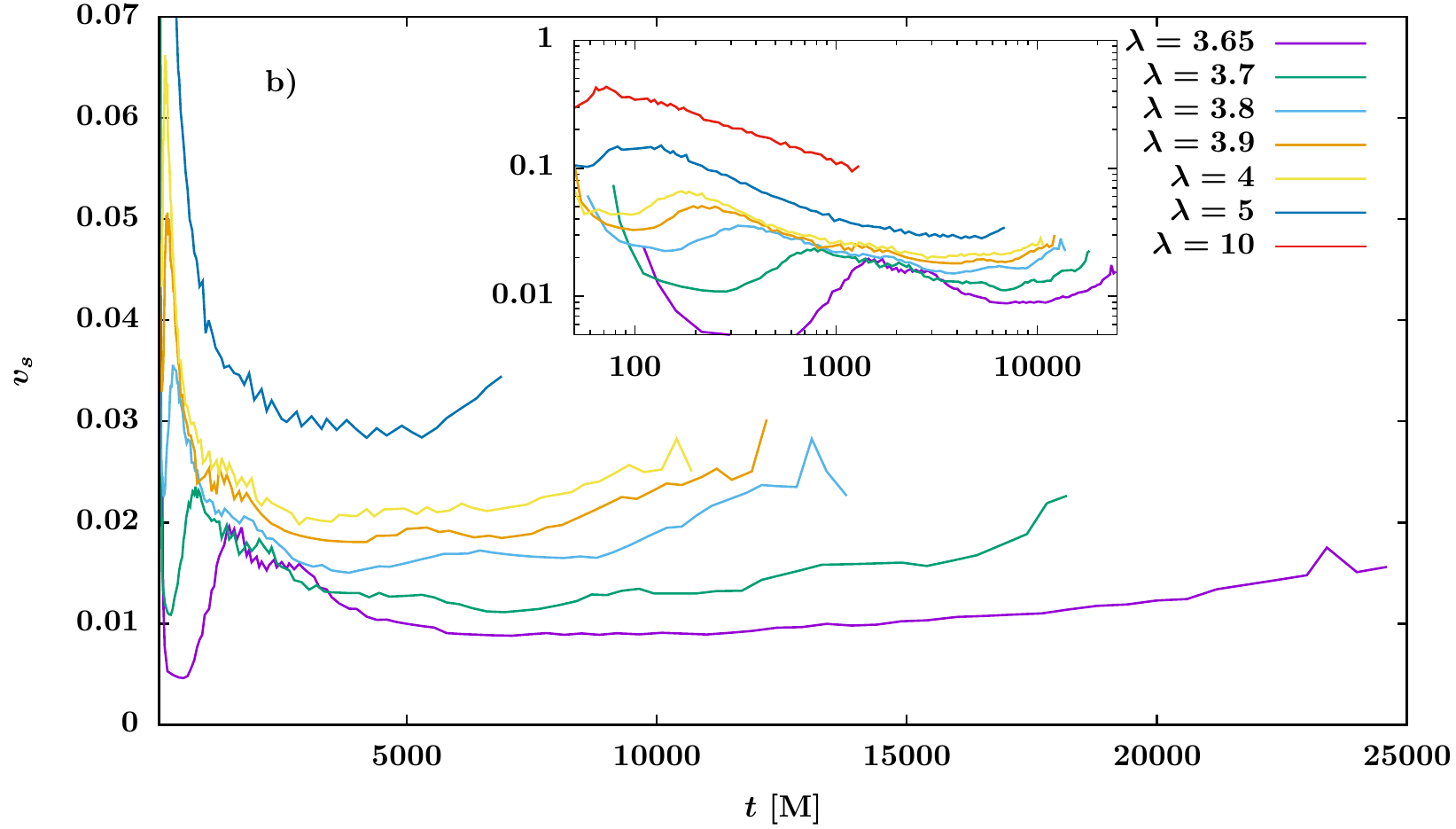}
  \includegraphics[width=0.48\textwidth]{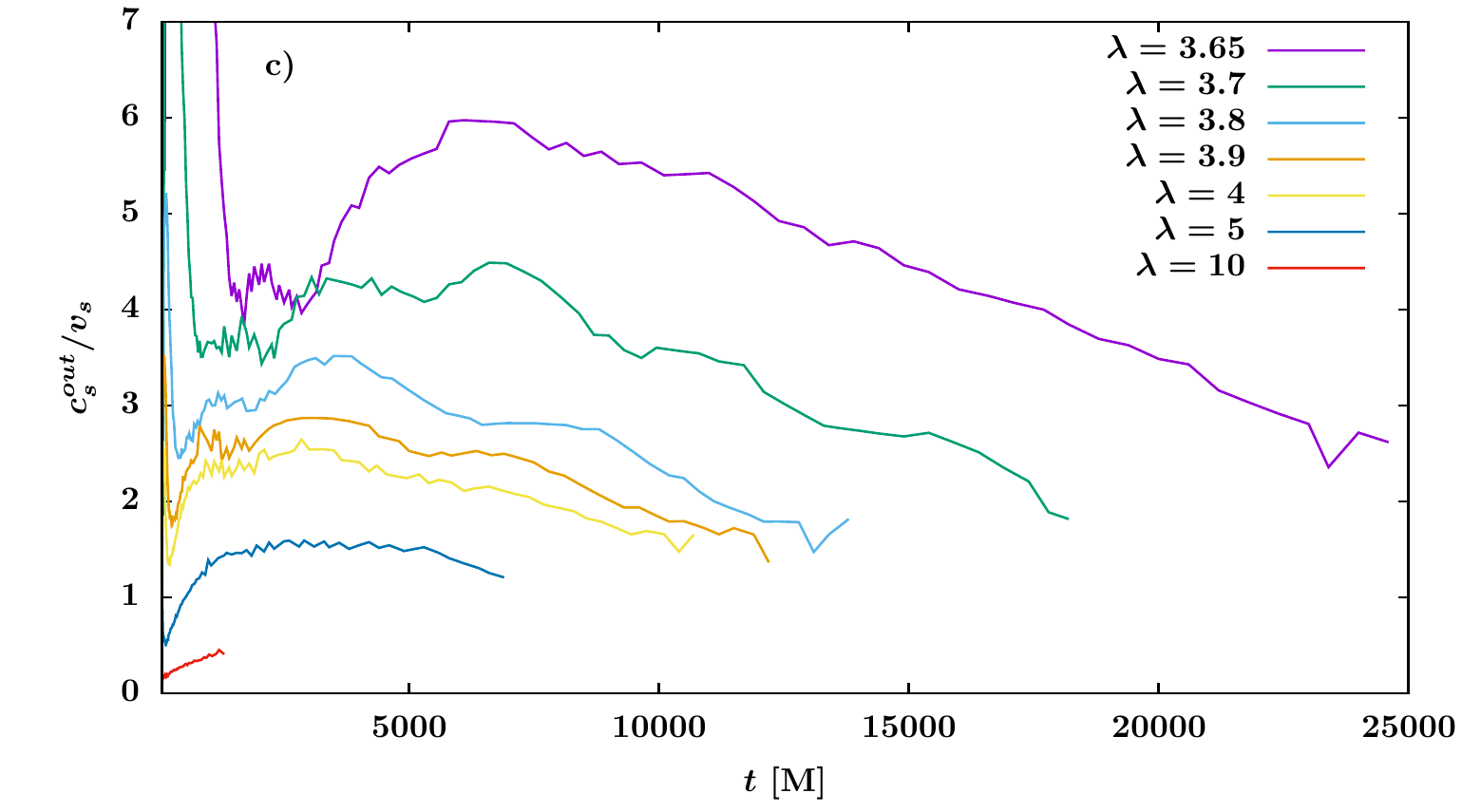}
 \caption{ Behaviour of the shock front for different values of $\lambda^{\rm eq}$ with $\mathcal{E}=0.0025$. Panel a) shows the position of the shock front, panel b) shows its velocity $v_s$ and panel c) displays the ratio of the preshock medium sound speed $c^{\rm out}_s$ to the shock front velocity $v_s$ during the evolution. \label{2D_Rs_position} }
\end{figure}

\subsection{2D computations} \label{results_2D}
After confirming the general outcomes of our previous PW study with full GR 1D simulations, we now continue with simulations in higher dimensions.
Because our initial data are rotationally symmetric (does not depend on $\phi$), is it possible to evolve only a slice spanning ($r,\theta$), and with constant $\phi$, thus assuming that the system will remain rotationally symmetric during the evolution. In order to justify to what extend and under which conditions this assumption is valid the comparison with 3D computation was needed. We will comment on that in Section~\ref{3D}.

However, within the 2.5-D approach, the $\phi$ component of the four-velocity is considered.  We prescribe this component in Boyer-Lindquist coordinates according to relations given by Eqs. (\ref{uphi}) and (\ref{uphi_null}).

We chose several exemplary simulations and we show their results in the form of snapshots.
Every snapshot contains four panels with the slices of $\mathfrak{M}$ and its equatorial profile, $\rho$ in arbitrary units, and $\lambda$ in geometrized units, and it is labelled by the time $t$ in geometrized units,  where $M$ is the mass of central black hole. The axes show the position in geometrized units.
The red colour in the slice of radial Mach number corresponds to the supersonic motion, blue regions indicate the subsonic accretion.
The shock front is located at the place, where the abrupt change from supersonic to subsonic motion occurs, hence it is represented with the white curve separating red region farther from the centre and blue region closer to the centre.
The sonic curves are also white, but they lie between the blue region farther and red region closer to the centre.
On top of the radial Mach number map the velocity streamlines are plotted in blue. These streamlines are computed from the velocity field at the given instant of time, hence they do not represent the actual history of a certain fluid parcel.

\begin{figure}
 \includegraphics[width=0.48\textwidth]{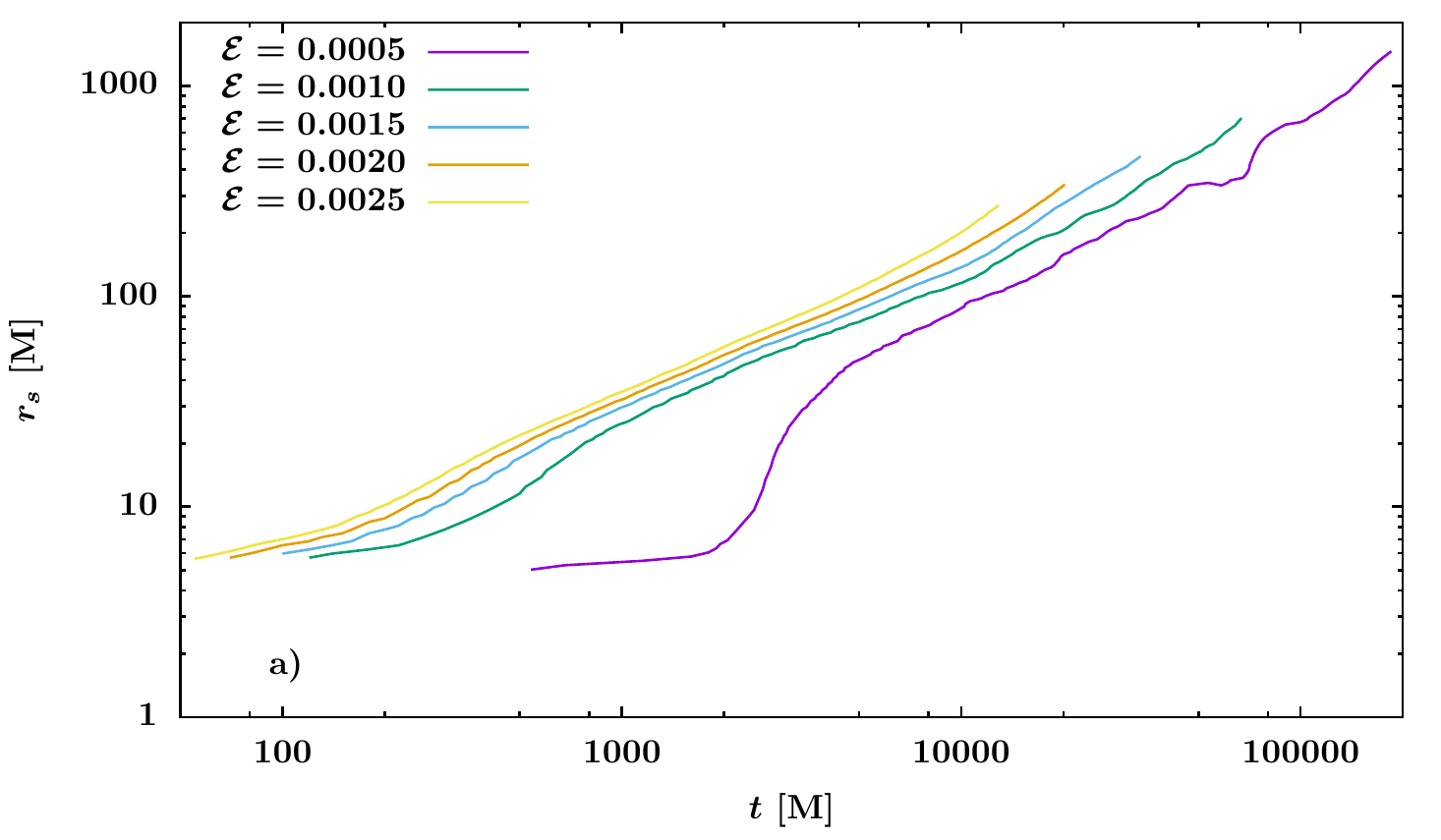}
 \includegraphics[width=0.48\textwidth]{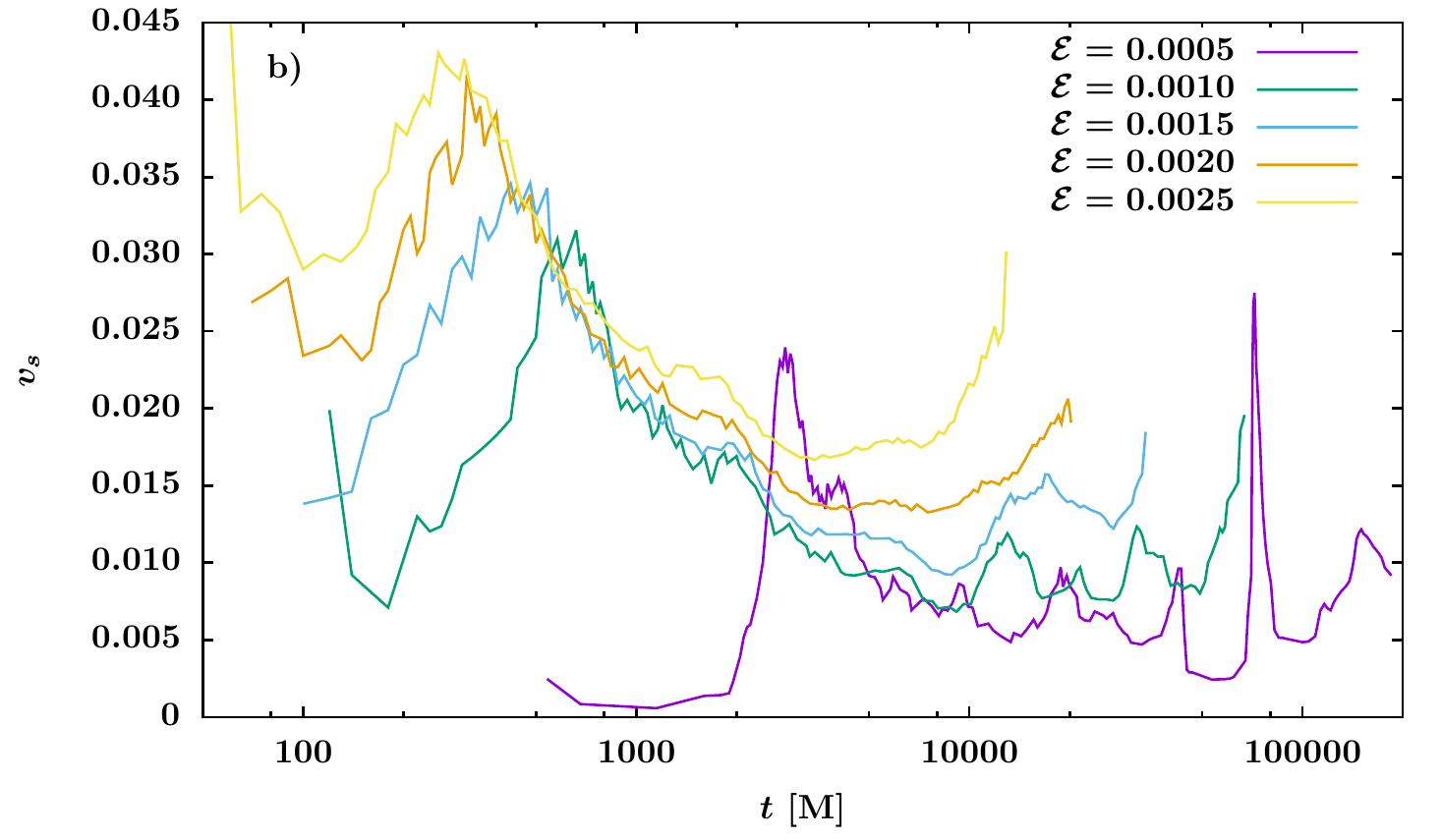}
 \includegraphics[width=0.48\textwidth]{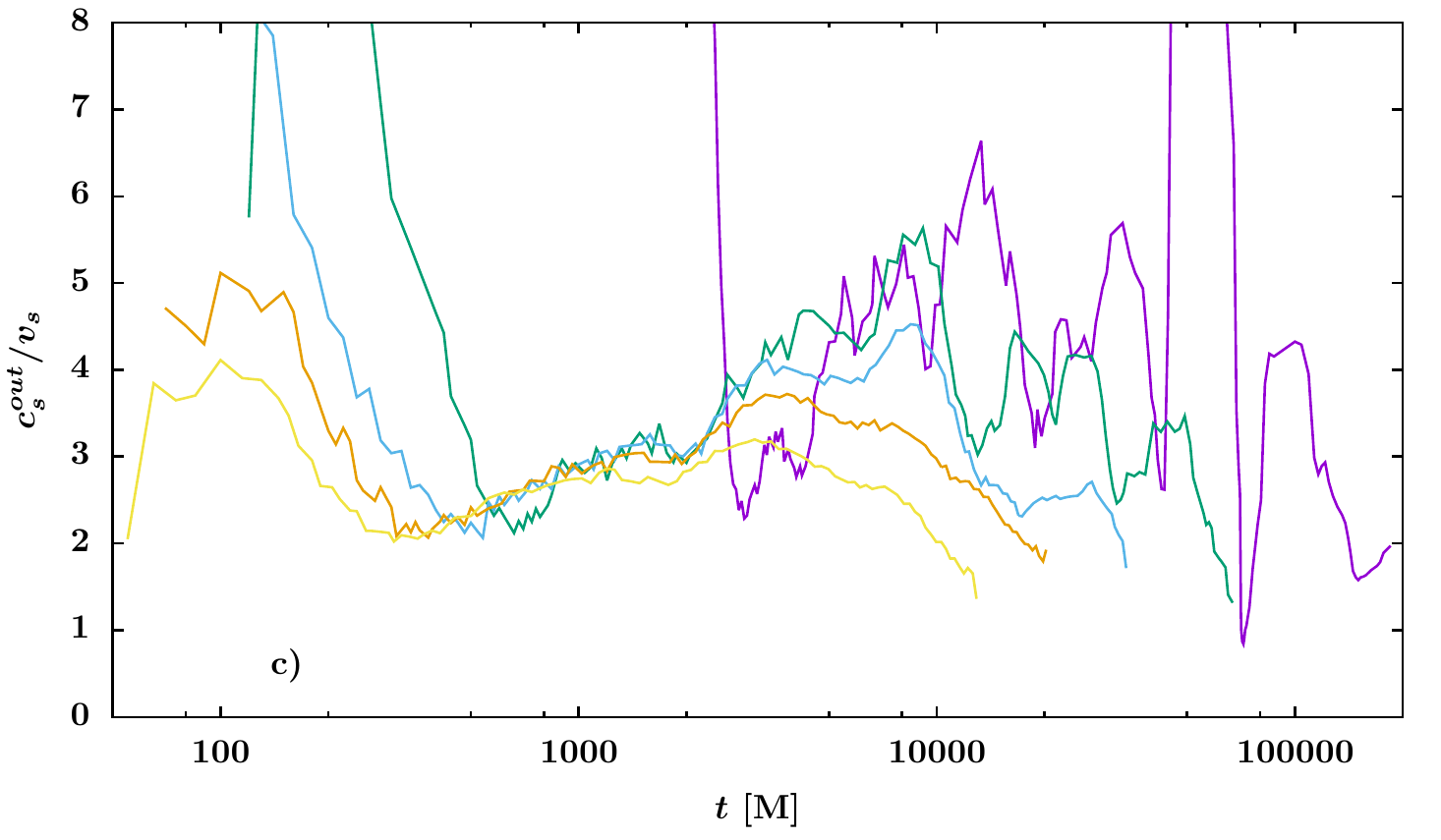}
 \caption{ The same as in Fig.~\ref{2D_Rs_position} for different values of $\mathcal{E}$ with $\lambda^{\rm eq}=3.85$ M. \label{2D_Rs_position_eps}
}
\end{figure}

\subsubsection{Bondi solution equipped with angular momentum}

At the beginning we check, that for pure Bondi solution with $\lambda=0$ we get the stationary solution of the flow with the properties consistent with the analytical solution.

After that we start to study the slowly rotating flows.
For the first simulation, we pick the parameters such that $\lambda^{\rm eq}$ belongs into the multicritical region (the region in the parameter space, where both the inner and outer sonic points exist in the 1D model), but it is very close to its upper bound, in particular
$\mathcal{E}=0.0025~M$, $\lambda^{\rm eq} = 3.6~M$, and $\gamma=4/3$. In that case, even when the shock solution is possible, it is not expected to appear, because the initial configuration is closer to the outer branch (there is no inner sonic point in the Bondi initial data).
In other words, we expect, that the rotation of the gas affects the profile of the Mach number in the innermost region in such a way, that it gets very close to 1, but does not touch it.

In Fig.~\ref{BondiRotInit}, the initial conditions at $t=0M$ are shown and the final state at $t_f=10^4M$ of the gas is depicted  in Fig~\ref{BondiRot}. At the later time, the simulation is already relaxed to the stationary state, which resembles the outer branch.
It is interesting to note, that for $t=t_f$ the Mach number actually crosses the $\mathfrak{M}=1$ line, however only on a very short radial range.
We would expect, that at the moment, when the inner sonic point appears, the shock creates and expands.
The reason, why it is not so in this simulation, is that the angular momentum is very close to the upper bound of the multicritical region, which is a boundary between the two distinct types of evolution.
In such case, the numerical evolution depends also on the resolution of the grid and other numerical settings.
In other words, if the parameters are close to such boundaries, then the simulations with different resolution can lead to a different type of evolution (i.e., the shock creates or not), hence it is difficult to find the exact critical value of the parameter.

This confirms our previous observation \citep{our_paper}, that in the multicritical region the evolution of the flow can tend to either the outer ``Bondi-like'' branch or to the shock solution, and the choice between these two possibilities is given by the initial conditions.
In particular, the profile of the Mach number in the innermost region and the presence of the inner sonic point is crucial for the shock development.

\begin{figure}
\begin{center}
 \includegraphics[width=0.5\textwidth]{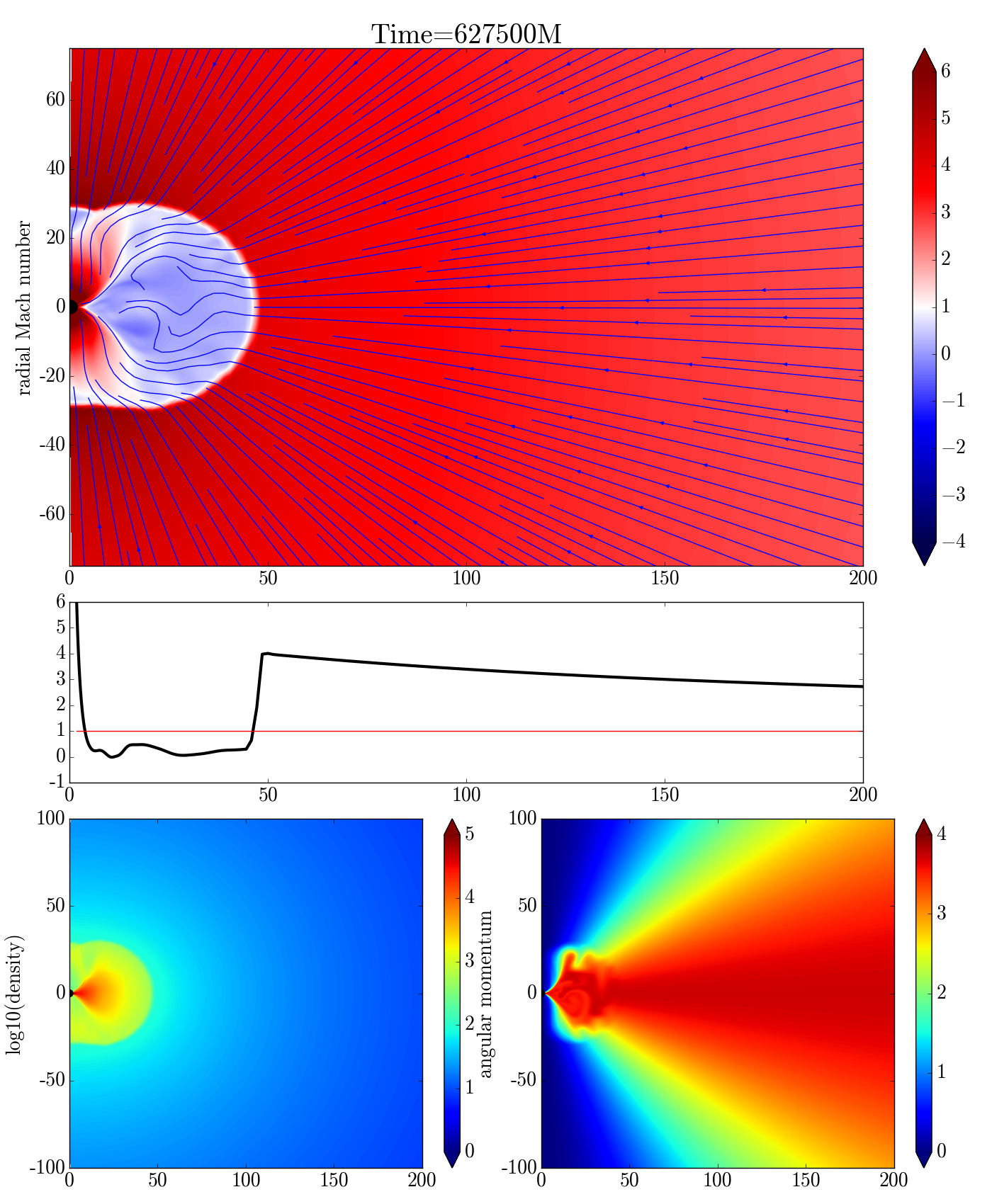}
 \includegraphics[width=0.5\textwidth]{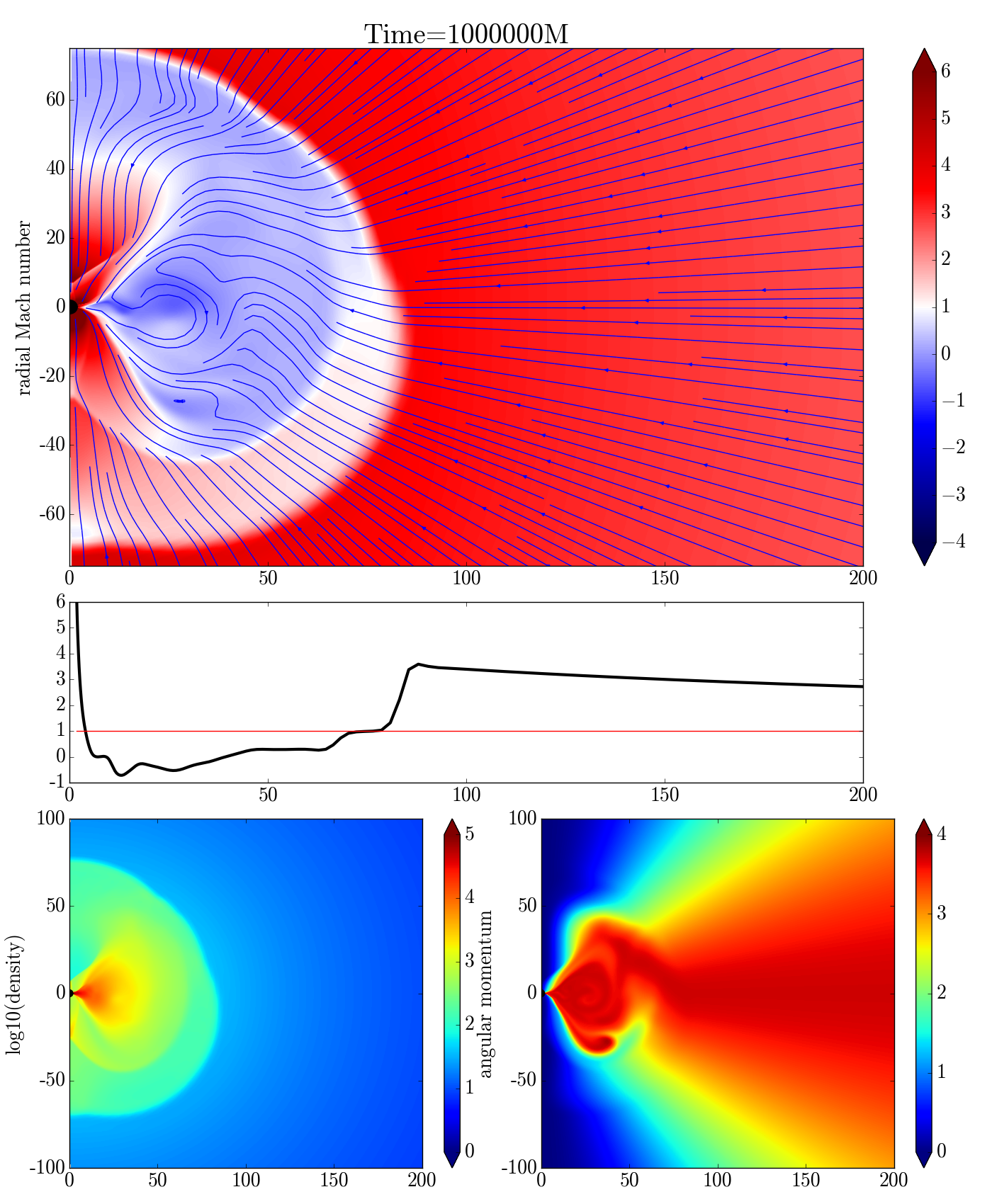}
  \caption{ Model \mI{}: Initial data with a shock with $\mathcal{E}=0.0005, \lambda^{\rm eq}=3.72$M. The shock bubble develops oscillations and changes size quasiperiodically, but does not acretes, nor does it expand outside $r_{\rm out}$.   \label{K112_Mach} }
 \end{center}
\end{figure}

In \cite{PTAproc} we showed a similar computation with $\lambda^{\rm eq}=3.79M, \mathcal{E}=0.001$, which is also close to the upper bound of the multicritical region.
Those computations were done in 3D with a different GR code, \texttt{Einstein toolkit}\footnote{\texttt{https://einsteintoolkit.org/}} \citep{ETtoolkit}.
The qualitative and quantitative agreement with the 1D solution is discussed there.
However the fact, that two very different codes working on different kind of grids (spherical grid logarithmic in radius versus Cartesian-like grid with the mesh refinement) show similar results is a good support for our results.

Last example is the case, when $\lambda^{\rm eq}$ is above the multicritical region, hence the outer type of solution is not possible. Therefore, the prescribed initial conditions are not close to a physical solution and the shock bubble has to appear. However, the position of the shock front is not stable and the shock is expanding until it meets the outer sonic point and the distant supersonic region dissolves, yielding only the inner type of accretion.

Two snapshots of the evolution in times $10^3M$ and $2\cdot 10^4M$ are given in Fig.~\ref{B10_II}.
On the first set of panels, the emergent shock bubble is seen, which is the expanding subsonic and more dense region.
Even though the shock front is thought to be very thin, in the numerical simulation it spans across several zones, which can be seen on the equatorial profile of $\mathfrak{M}$.
In the last snapshot, the outer supersonic region is already dissolved and only a small supersonic region very close to the black hole can be found.

The time dependence of the shock and sonic point positions in the equatorial plane is shown in Fig.~\ref{B10_II_rs}, where it can be seen, that the shock meets the outer sonic point at about $t\sim15000~M$, which corresponds to about 0.75~s in case of a typical microquasar with $M=10~M_\odot$. This is way too short in comparison with the observational data concerning the quasiperiodic oscillations, which change frequency during few weeks. 
Hence, if we want to use an oscillatory shock front model to explain the observed low-frequency QPOs, we need to obtain a solution with long lasting shock front oscillating around the slowly changing mean position, and not the fast expanding shock bubble like  was observed in this case. 
The first step is to find an existing stationary solution with a standing shock.
As we already mentioned, such solution is not expected to occur, if we start the evolution with initial conditions close to the Bondi solution.

We can be, however, interested in the dependence of the shock propagation on the properties of the gas.
On that account, we performed two sets of simulations, where we changed the angular momentum and the energy of the flow such that they are above the multicritical solution, and we followed the expansion of the shock front until it met the outer sonic point.

In Fig.~\ref{2D_Rs_position} we show the behaviour of the shock front during the evolution for several different values of $\lambda^{\rm eq}$. For  $\lambda^{\rm eq}=3.65$ M, just above the multicritical region, the growth of the shock bubble is slow with a long transient period, during which it is unclear, if the bubble converges to the stationary state, or it rather expands outwards.
With increasing $\lambda^{\rm eq}$ the bubble expands faster.
The velocity of the shock front ranges between $0.005c$ up to $0.4c$ for the highest angular momentum.
Panel c) shows the ratio of the preshock medium sound speed $c^{\rm out}_s$ to the shock front velocity $v_s$. Except for the highest $\lambda^{\rm eq}=10$ M, the shock front expands by a velocity, which is a few times lower than the corresponding preshock sound speed, for the lowest $\lambda^{\rm eq}=3.65$ M the ratio $c^{\rm out}_s/v_s$  reaches the value up to 6.

In Fig.~\ref{2D_Rs_position_eps} we present similar results obtained for different values of $\mathcal{E}$.
Here, the position of the outer sonic point differs significantly for different $\mathcal{E}$, hence also the time for the shock front to reach the outer sonic point varies considerably.
The plots are therefore given in logarithmic scale of time.

\begin{figure}
\begin{center}
 \includegraphics[width=0.5\textwidth]{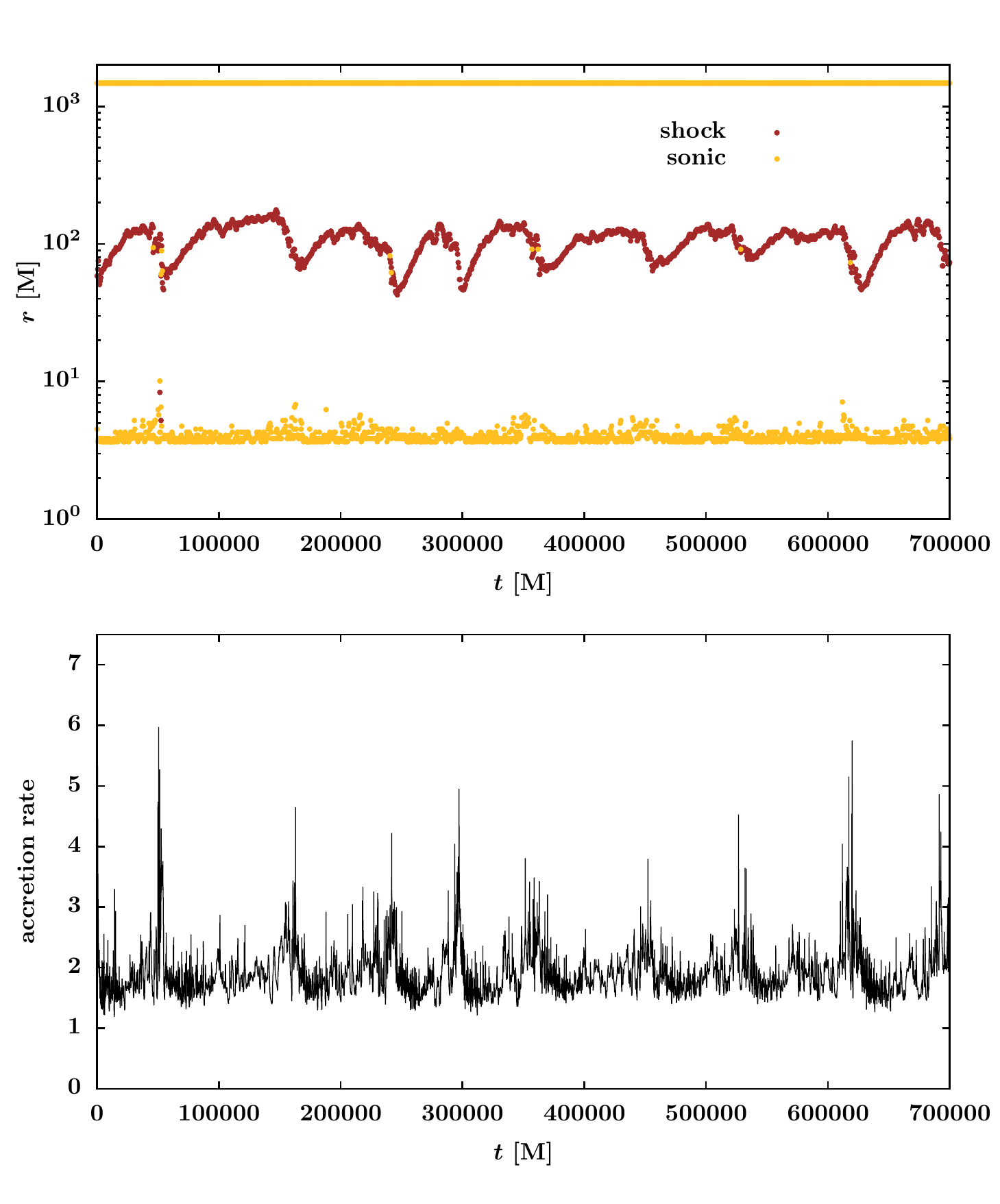}
 \caption{ Model \mI{}: The shock and sonic point position and the corresponding mass accretion rate through the inner boundary.    \label{K112_rs} }
 \end{center}
\end{figure}

\begin{figure}	
 \includegraphics[width=0.5\textwidth]{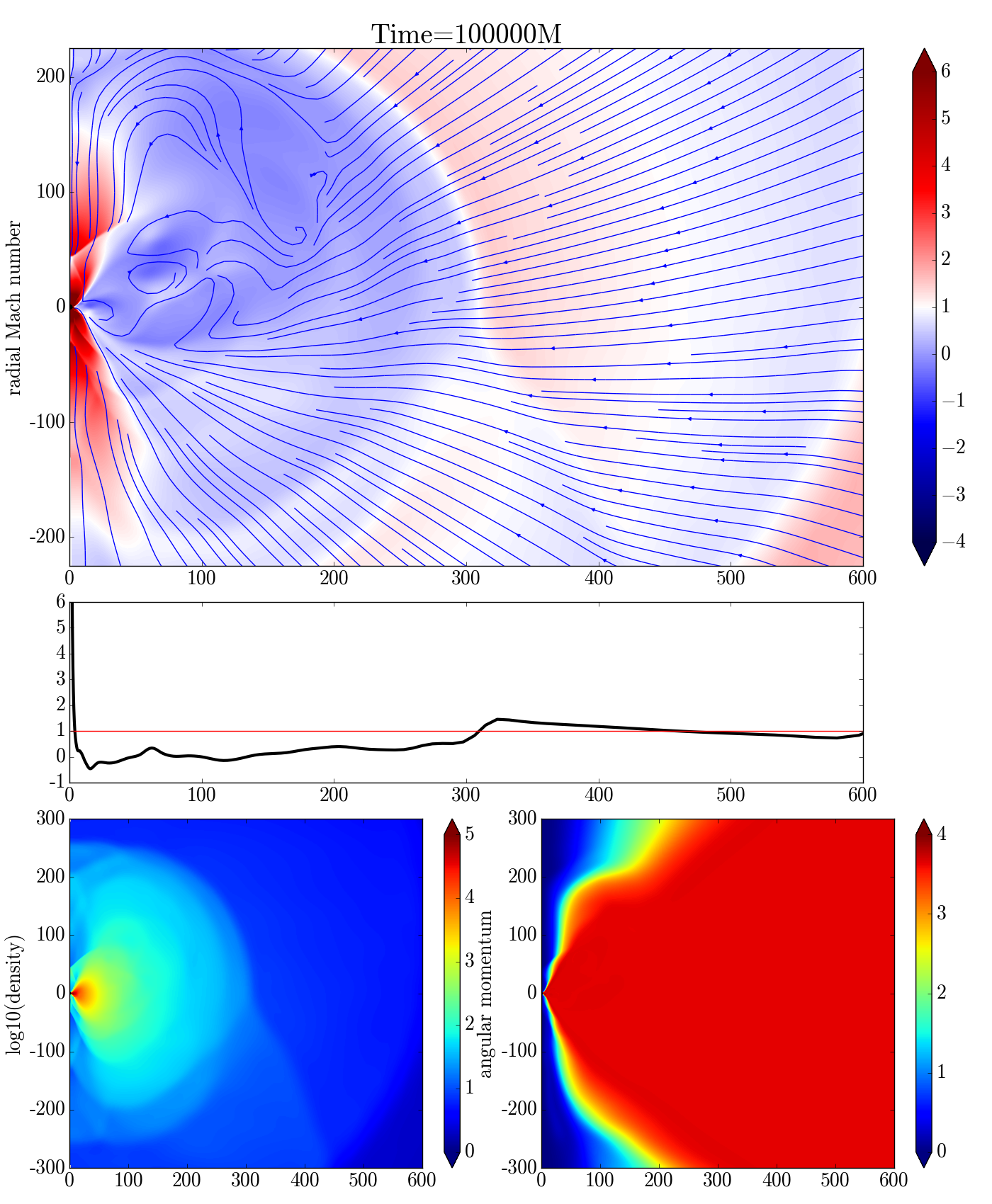}
 \includegraphics[width=0.5\textwidth]{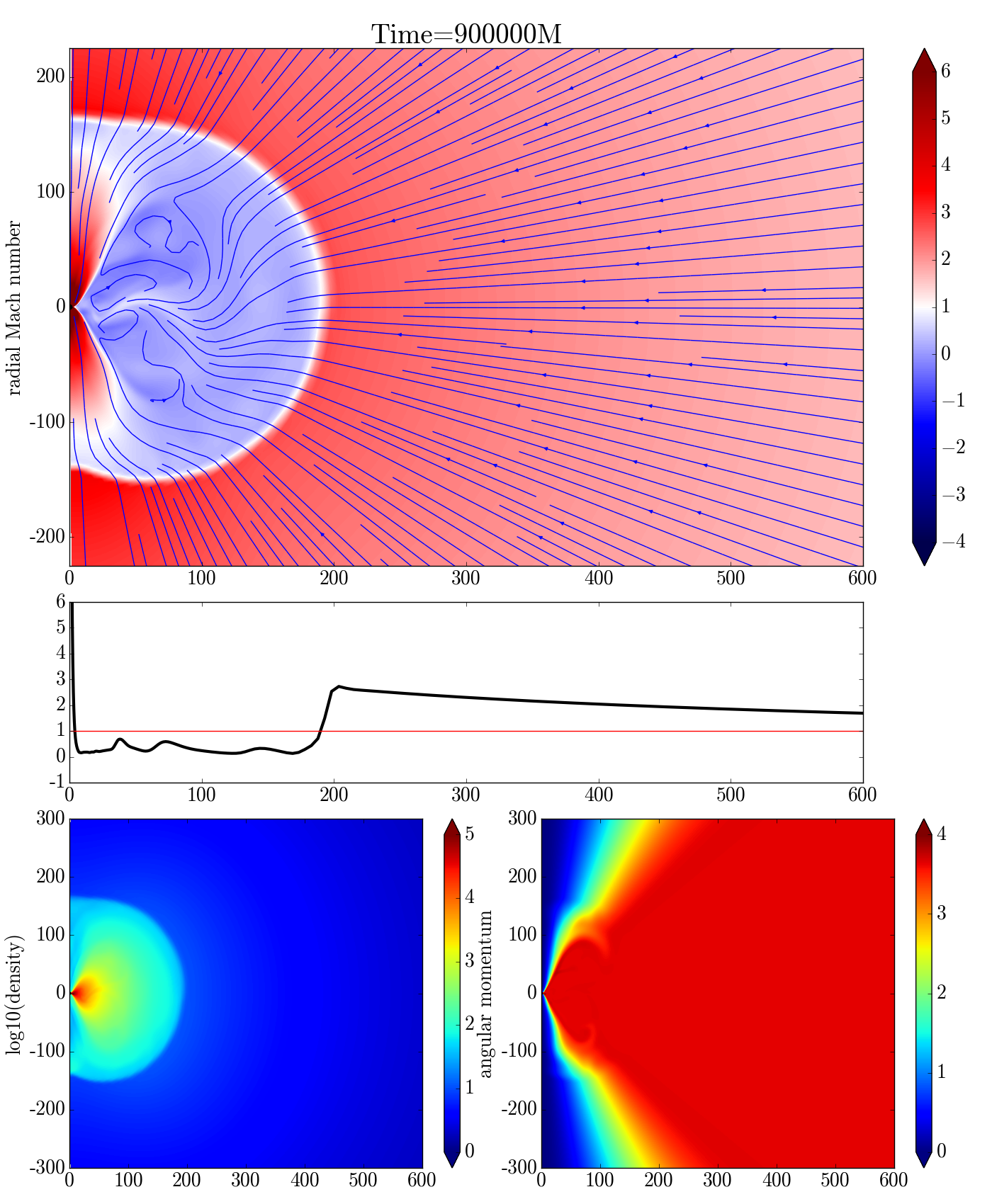}
   \caption{ Model \mHc{}: Initial data with a shock and constant angular momentum distribution ($\mathcal{E}=0.0005, \lambda=3.65$M) in the cone with the half-angle $\theta_c=\pi/4$.   \label{K430_Mach} }
\end{figure}

The velocity of the shock front decreases with the decreasing energy, and the same holds for its ratio to the sound speed of the preshock medium.
The trend is similar as for decreasing angular momentum.
The reason is, that for decreasing energy, the multicritical region exists for higher value of angular momentum, as we have seen earlier.
Hence, when we decrease the energy and keep the angular momentum constant, we are approaching the multicritical region in the parameter space.
That is well documented on the case with the lowest energy $\mathcal{E}=0.0005$ (the purple lines in Fig.~\ref{2D_Rs_position_eps}), which lies very close the multicritical region boundary and so the shock bubble waits for a very long time before it starts to expand.
Therefore, we conclude, that the shock front velocity depends on the distance of our chosen parameters ($\lambda, \mathcal{E}$) from the multicritical region in the parameter space and that it is typically a few times lower than the sound speed in the preshock medium.

\subsubsection{Shock solution}
Because we want to find out, if there are any stationary or oscillating solutions with shocks also in 2D, similarly as in 1D, we have to prescribe the initial conditions, which are closer to the shock solution branch than the Bondi-like solution.
These initial conditions are described in Section~\ref{Ini_shock}.

Inspired by the range of shock solution in the PW case, we performed a set of simulations with changing angular momentum $\lambda^{\rm eq} \in [3.52M,3.7M]$,
while keeping $\mathcal{E}=0.0025$ and $\gamma=4/3$. (Note that the range in 1D paper was defined for $\mathcal{E}=0.0001$, not $0.0025$. Here, the range is such, that for the lowest $\lambda$ the shock bubble accretes and for the higher values it expands, so it covers the whole interesting interval.)
These simulations showed, that for lower angular momentum the stationary state is obtained, while for higher angular momentum the shock bubble again expands and merges with the outer sonic surface, which is located around 280M.
The details of these computations are given in \cite{IAUS_proc}. 

When we choose lower value of energy, the outer sonic surface is located further, so that there is more space, where the shock existence is possible. We performed several simulations with  $\mathcal{E}=0.0005$ and $\mathcal{E}=0.0001$, for which $r_{\rm out}=1475~M$ and $r_{\rm out}=7486~M$, respectively.

\begin{figure}
 \includegraphics[width=0.5\textwidth]{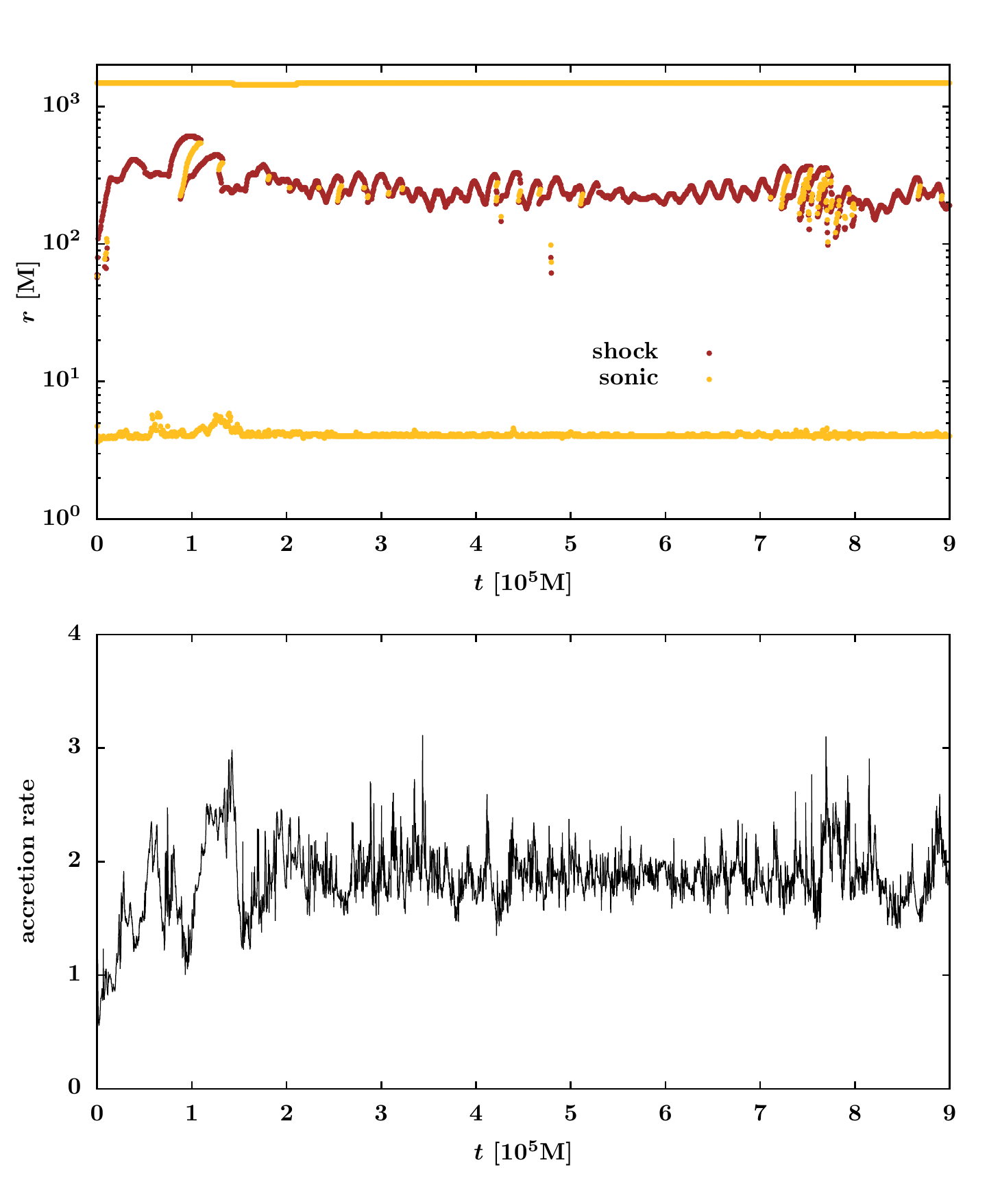}	
 \caption{Model \mHc{}:  The shock and sonic point position and the corresponding mass accretion rate through the inner boundary.     \label{K430_rs} }
\end{figure}

For higher values of angular momentum we found the shock bubble unstable - the eddies emerge in the flow, the bubble is growing for some time, after which a quick accretion occurs accompanied by shrinking of the bubble, which is also oscillating in vertical direction. However, it does not expand or accrete completely.
Several snapshots of the evolution are given in Fig.~\ref{K112_Mach}, where the shock bubble has different shape and size at different times and the asymmetry with respect to equator can be seen.
The time dependence of the shock front position in the equatorial plane is given in Fig.~\ref{K112_rs}.

\begin{table*}
\begin{tabular}{c|ccccccccccc}
 & $\mathcal{E}$&$\lambda^{\rm eq}$ [M]& $a$ &IC & res & shock & $r_s$ [M] & $\bar{r}_s$ [M] &  f [M$^{-1}$] & Fig \\ \hline
\mD{} & 0.0025 & 3.6   & 0 &  Bondi & 256 x 128 &no & -- & -- & & \ref{BondiRot} \\ %
\mE & 0.0025 & 3.8   & 0 &  Bondi & 256 x 128 &EX & 5 -- 282 & -- & & \ref{B10_II}, \ref{B10_II_rs} \\ 
\mF & 0.0005 & 3.58   & 0 & 1D+sph & 384 x 192 & AC & --  & -- & \\ 
\mG  & 0.0005 & 3.59   & 0 & 1D+sph & 384 x 192 & OS & 28 -- 41  & 35 & $\sim 3\cdot 10^{-3}$ & \\ 
\mH & 0.0005 & 3.6   & 0 & 1D+sph & 384 x 192 & OS & 31 -- 45  & 39 & $\sim 5\cdot 10^{-3}$ & \\ 
\mI & 0.0005 & 3.72 & 0 & 1D+sph & 384 x 192 & OS & 29 -- 177  & 106 & $\sim 1\cdot 10^{-5}$ & \ref{K112_Mach}, \ref{K112_rs} \\ 
\mJ & 0.0005 & 3.78  & 0 & 1D+sph & 384 x 192 & OS & 10 -- 521  & 216 & $\sim 4\cdot 10^{-6}$ & \\ 
\mK & 0.0001 & 3.6  & 0 & 1D+sph & 384 x 192 & AC & --  & --  & \\ 
\mL & 0.0001 & 3.72  & 0 & 1D+sph & 384 x 192 & OS & 26 -- 62  & 43 & $\sim 6\cdot 10^{-5}$ & \\ 
\mM & 0.0001 & 3.78  & 0 & 1D+sph & 384 x 192 & OS & 21 -- 146  & 70 & $\sim 2.2\cdot 10^{-5}$ & \\ 
\mN & 0.0001 & 3.82   & 0 & 1D+sph & 384 x 192 & OS & 10 -- 226  & 86 & $\sim 1.7\cdot 10^{-5}$ & \\ 
\mO & 0.0001 & 3.86  & 0 & 1D+sph & 384 x 192 & OS & 10 -- 359  & 124 & $\sim 1.9\cdot 10^{-5}$ & \\ 
\mEa & 0.0005 & 3.34  & 0.3 & 1D+spin & 384 x 192 & AC & --  & -- & \\
\mEb & 0.0005 & 3.35  & 0.3 & 1D+spin & 384 x 192 & OS & 18 -- 36  & 30 & $\sim 7\cdot 10^{-3}$ & \\
\mEc & 0.0005 & 3.40  & 0.3 & 1D+spin & 384 x 192 & OS & 32 -- 81  & 59 & $\sim 2.3\cdot 10^{-3}$ & \\
\mEd & 0.0005 & 3.48  & 0.3 & 1D+spin & 384 x 192 & OS & 26 -- 200  & 100 & $\sim 2\cdot 10^{-5}$ & \\
\mEe & 0.0005 & 3.49  & 0.3 & 1D+spin & 384 x 192 & EX & --  & -- & \\
\mP & 0.0005 & 2.7  & 0.8 & 1D+spin & 384 x 256 & AC & --  & -- & \\ 
\mQ & 0.0005 & 2.8  & 0.8 & 1D+spin & 384 x 256 & OS & 13 -- 80  & 39 & $\sim 5.5\cdot 10^{-5}$ & \ref{K126_UHR},\ref{K126_UHR_rs}  \\ 
\mR & 0.0005 & 2.9  & 0.8 & 1D+spin & 384 x 256 & EX & 100 -- 1450  &  & \\ 
\mGa & 0.0005 & 2.4 & 0.95 & 1D+spin& 576 x 192 & OS & 10 -- 58 & 29 & $\sim 7\cdot 10^{-5}$ & \\ 
\mGb & 0.0005 & 2.42 & 0.95 & 1D+spin& 576 x 192 & OS & 11 -- 74 & 31 & $\sim 8\cdot 10^{-5}$  & \ref{K680_rs} \\ 
\mHa & 0.0005 & 3.5 & 0.0 & 1D+cone& 384 x 192 & AC & -- & --& & \\ 
\mHb & 0.0005 & 3.6 & 0.0 & 1D+cone& 384 x 256 & OS & 110 -- 134 & 121 & $\sim 2.8\cdot 10^{-5}$ & & \\ 
\mHc & 0.0005 & 3.65 & 0.0 & 1D+cone& 384 x 256 & OS & 98 -- 366  & 235 & $\sim 8\cdot 10^{-6}$ & \ref{K430_Mach},\ref{K430_rs}& \\ 
\mHd & 0.0005 & 3.72 & 0.0 & 1D+cone& 384 x 192 & EX & 90 -- 1477 & --& & \\ 
\mIa & 0.0005 & 3.65 & 0.0 & 1D+sph& 256 x 128 x 96 & OS &  68 -- 81 & 76 & & \ref{K500},\ref{K500_rs}\\ 
\end{tabular}
\caption{Summary of computed 2D models. All models have $\gamma=4/3$. Initial conditions (IC) are described in Sections \ref{Ini_Bondi} (``Bondi''), \ref{Ini_shock_sph} (``1D+sph''), \ref{Ini_shock_cone} (``1D+cone'') or \ref{Ini_spin} (``1D+spin''). In case of 1D+cone simulations $\theta_c=\frac{\pi}{4}$. The shock behaviour is labeled as ``no'' for solution without a shock, ``EX'' for solution with expanding shock, which merges with the outer sonic point, ``OS'' for solutions with oscillating shock front and ``AC'' for solutions, in which the shock front is accreted almost immediately. The next three columns show the range, in which the shock is moving, the average position of the shock and the frequency of the oscillation and flares in the mass accretion rate through the inner boundary, if there are any. The last column contains the reference to the corresponding Figures.  \label{t:2D-models} }
\end{table*}

\begin{figure}	
\begin{center}
 \includegraphics[width=0.5\textwidth]{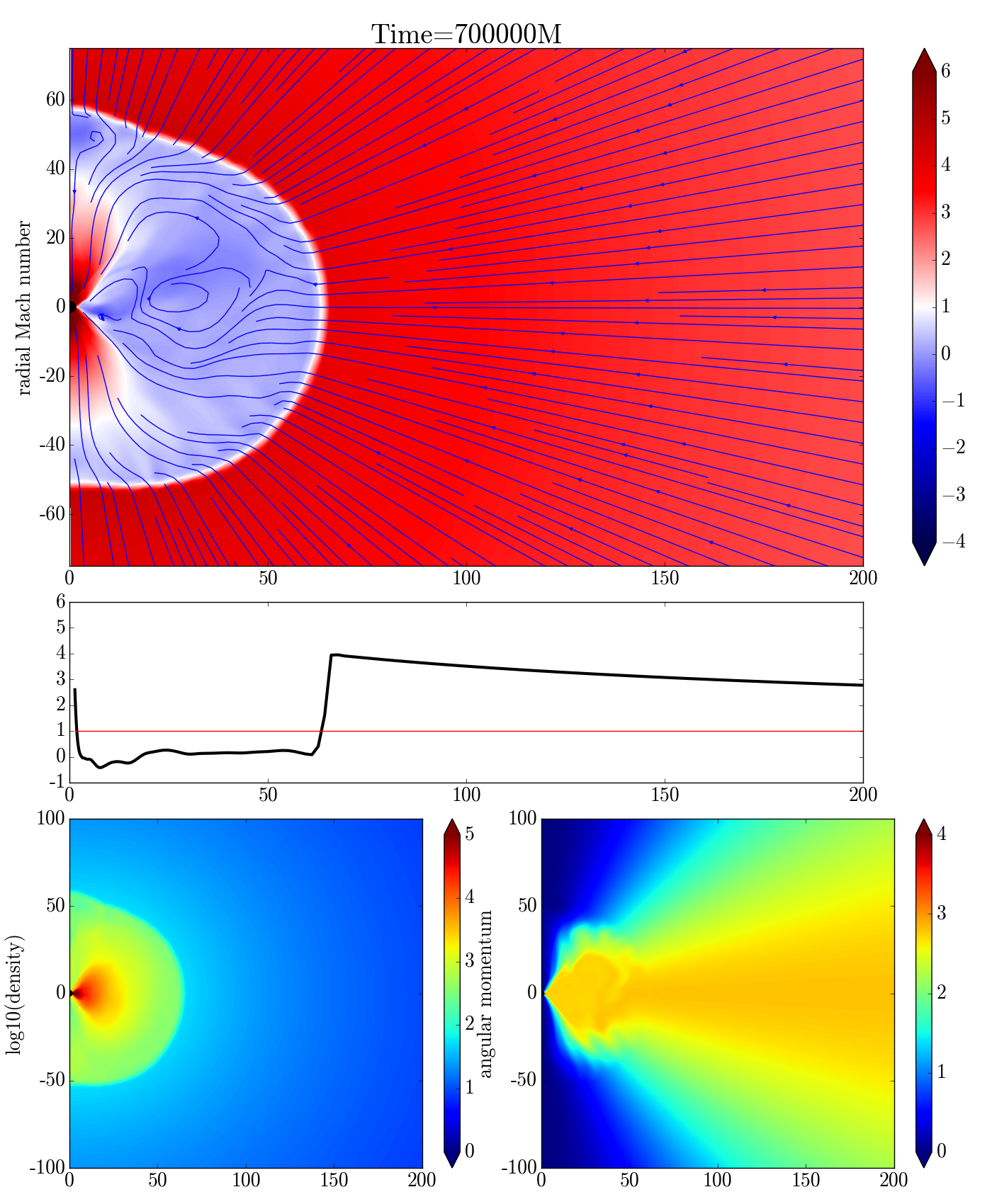}
 \caption{ Model \mQ{}: moderately spining black hole ($\mathcal{E}=0.0005, a=0.8, \lambda^{\rm eq}=2.8$M) with oscillating shock.   \label{K126_UHR} }
 \end{center}
\end{figure}

This repetitive process causes also flaring in the mass accretion rate through the inner boundary, see Fig~\ref{K112_rs}. For $10~M_\odot$ black hole, $t=10^6~M$ corresponds to approx. 50~s, hence the flares occur on a similar time scale as in some of the flaring states of microquasars,  e.g. the heartbeat state of GRS 1915+105 \citep{2000A&A...355..271B} or IGR J17091-3624 \citep{2012ApJ...747L...4A}. Moreover, we have shown, that the sources in this state show the evidence of nonlinear mechanism behind the emission of the hard component, which corresponds to the low angular momentum flow  \citep{nas_nonlin,nas_chaos_small}. However, this fact is only an indirect evidence for the presence of the shock, because there are other possible explanations of the flares, e.g. they can be a result of the radiation pressure instability in the disc \citep{2015A&A...574A..92J,2016arXiv160909322G}.

Another type of simulations are those with different distribution of angular momentum (see Section \ref{Ini_shock_cone}). Three snapshots of such evolution are given in Fig.~\ref{K430_Mach} and the corresponding shock position and mass accretion rate is given in Fig.~\ref{K430_rs}.

We found qualitatively similar behavior for different values of parameters.
The results are summarized in Table~\ref{t:2D-models}. For $\mathcal{E}=0.0005$ we identified the interval of angular momentum values for which there is oscillating shock to be $[3.59{\rm M},\,3.78{\rm M}]$ (models \mG{}-\mJ{}). For $\mathcal{E}=0.0001$ the interval is $\lambda\in[3.72{\rm M},\,3.86{\rm M}]$ (models \mL{}-\mO{}). In both series of simulations it is clearly visible, that increasing the value of angular momentum within the interval causes increase of the amplitude of shock oscillations. So oscillations appear sharpest just before the upper limit of the angular momentum range and the oscillation frequency peaks in the spectrum are clearly visible in those cases.

We can compare the series of simulations \mF{}-\mJ{} with \mHa{}-\mHd{} which have the same values of parameters, but differ in distribution of angular momentum (modulation by $\sin^2\theta$ versus constant angular momentum in a cone). It is clear that the results are very similar: in both cases there is a range of angular momentum in which the shock oscillating with comparable frequencies is present, with amplitude of oscillations increasing with angular momentum value. This shows, that the main conclusions from our simulations are resistant to changes in spatial distribution of rotation profile.

\subsubsection{Rotating black hole}
We chose three values of the spin $a$, which can represent quite well the estimated range of spins of the know microquasars, in particular $a=0.3$ (which could be a representative value for XTE J1550-564, H1743-322, LMC X-3, A0620-00), $a=0.8$ (M33 X-7, 4U 1543-47, GRO J1655-40) and $a=0.95$ (Cyg X-1, LMC X-1, GRS 1915+105).
The estimated values of the spin are taken from \cite{McClintock2014}.

For $a=0.3$ and $\lambda^{\rm eq}_g \in[3.35~\mathrm{M},3.48~\mathrm{M}]$ we observed a long lasting oscillating shock front. For $\lambda^{\rm eq}_g=3.34~\mathrm{M}$ (and lower values) the shock is accreted, and for $\lambda^{\rm eq}_g=3.49~\mathrm{M}$ (and higher values) the shock surface expands. The Table~\ref{t:2D-models} contains details about five selected simulations of $a=0.3$ scenarios: values on both sides of left and right boundary of interval and one in the middle of the interval of $\lambda^{\rm eq}_g$ values corresponding to stable shock existence.

\begin{figure}
\begin{center}
 \includegraphics[width=0.5\textwidth]{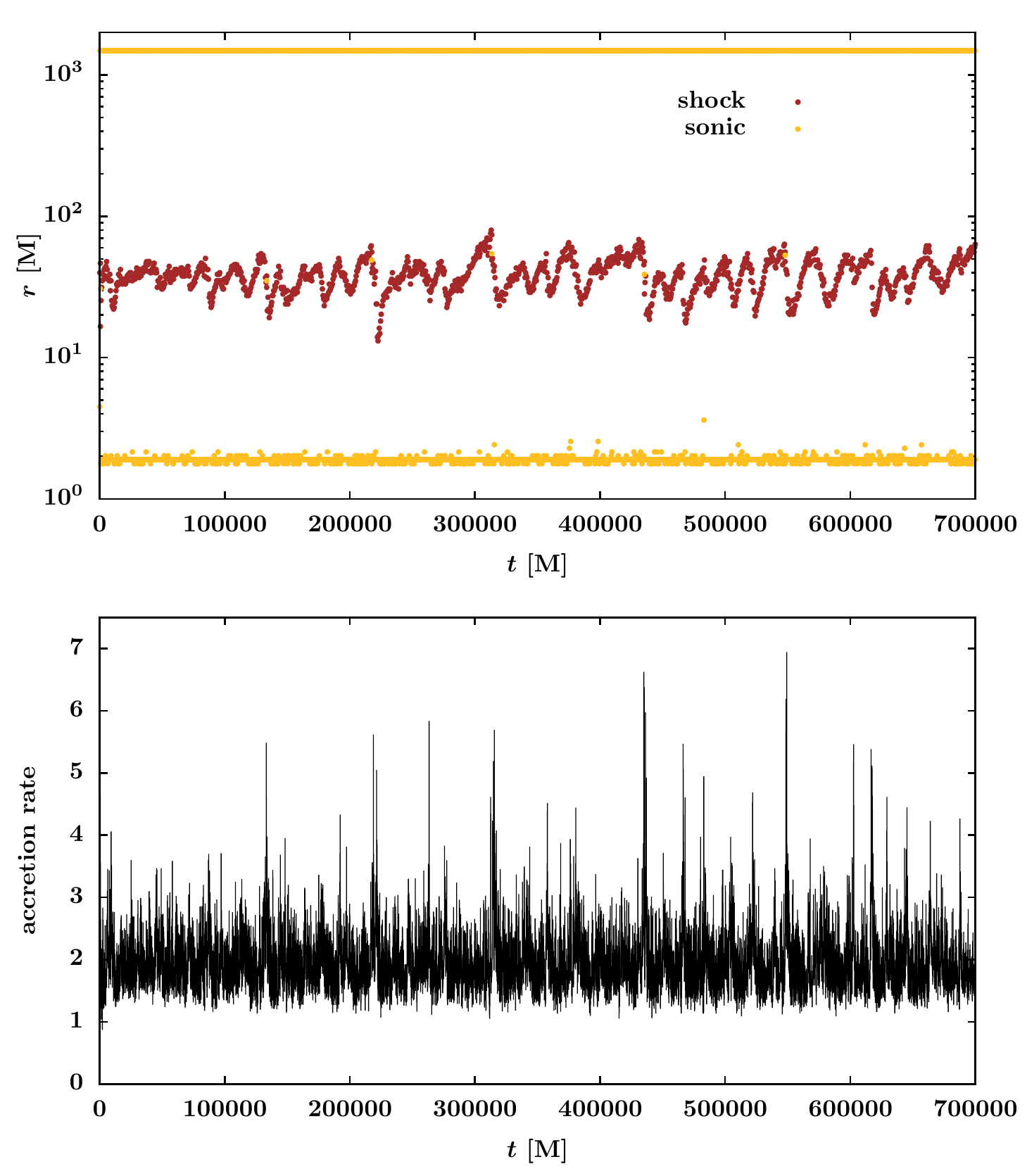}
 \end{center}
 \caption{Model \mQ{}: moderately spining black hole ($\mathcal{E}=0.0005, a=0.8, \lambda^{\rm eq}=2.8$M) with oscillating shock.  \label{K126_UHR_rs} }
\end{figure}

Next in the Table~\ref{t:2D-models} the results of several simulations with $a=0.8$ are shown. In this case the interval of $\lambda^{\rm eq}_g$ for which there are solutions with oscillating shock appears in the range of lower values than for $a=0.3$.
For  $\lambda^{\rm eq}_g = 2.7$~M (model \mP{}), the shock is accreted, for $\lambda^{\rm eq}_g = 2.8$~M (model \mQ{}) the accretion flow contains a long lasting oscillating shock front, and for $\lambda^{\rm eq}_g = 2.9$~M (model \mR{}) the shock bubble is expanding.
The qualitative behavior of the shock front is very similar like in the Schwarzschild case, only the exact values of parameters (mainly the angular momentum of the flow) differs. Again increase of the value of angular momentum within the oscillating shock interval corresponds to increase of the amplitude of oscillations. Series of models \mEb{}-\mEd{} is very similar in this regard to series \mG{}-\mJ{} and \mL{}-\mO{}.

For higher values of spin we found the computations to be more sensitive on the resolution, mainly in the radial direction close to the black hole. If the resolution is not sufficient, then the the shock bubble can accrete even after a long oscillatory evolution. However when we increase the resolution for the initial conditions, the shock persists in the evolution, even when it comes very close to the black hole.
Therefore we increased the radial resolution for $a=0.95$ to $N_{\rm r}=576$. Even with this high resolution, the shock bubbles in simulations with $\lambda^{\rm eq}=2.42$~M and $\lambda^{\rm eq}=2.45$~M are accreted after long evolution with repeated shock front oscillations.

We conclude that the qualitative results obtained in the non-spinning case can be generalized for spinning black holes. What can be clearly seen from the simulations, is that
the behavior similar to the case of nonrotating BH is observed for much lower values of the angular momentum, so the spin of the black hole ``adds'' to the rotation of the gas.
Increasing the value of the spin of black hole decreases both limiting values of angular momentum of accreted gas and leads to a narrower interval in which the oscillating shock solutions are observed.

\begin{figure}
\begin{center}
 \includegraphics[width=0.5\textwidth]{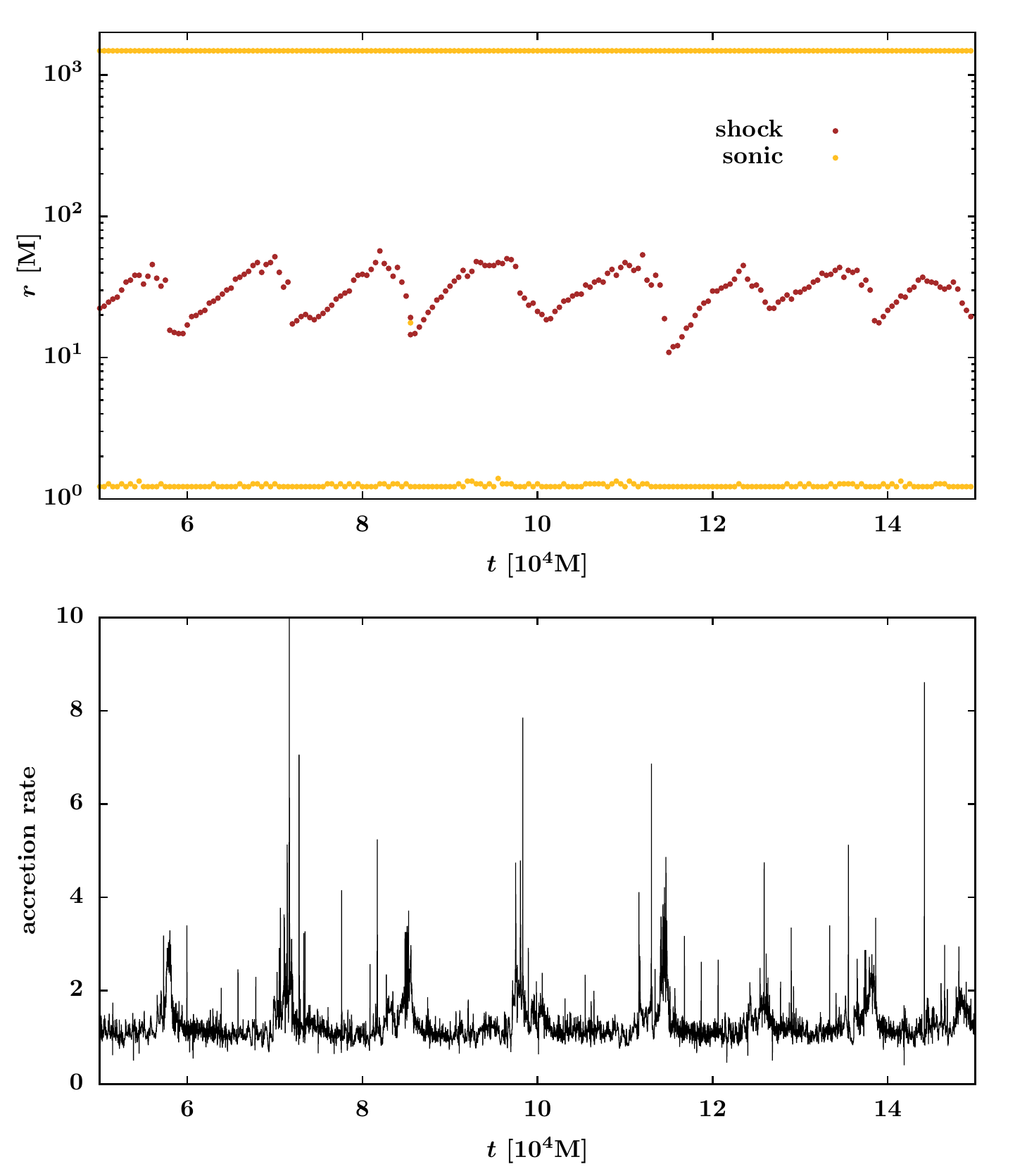}
 \end{center}
 \caption{Model \mGb{}: rapidly spinning black hole ($\mathcal{E}=0.0005, a=0.95, \lambda^{\rm eq}=2.42$M) with oscillating shock.  \label{K680_rs} }
\end{figure}

\subsection{3D computations} \label{3D}
\begin{figure}
\begin{center}
 \includegraphics[width=0.5\textwidth]{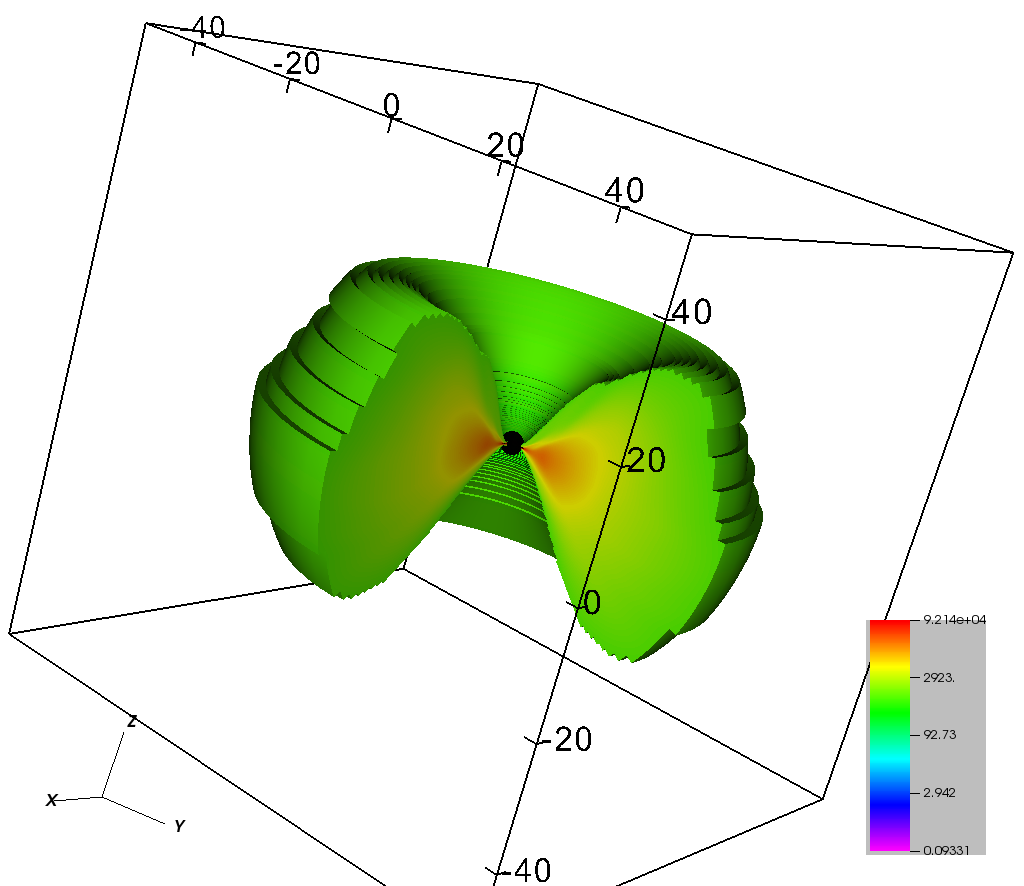}
 \end{center}
 \caption{Model \mIa{}: Density distribution in the three-dimensional simulation at time 28000M, within the innermost 50 gravitational radii from the black hole. Resolution of this run is 384 x 192 x 64. The colour scale and threshold is adopted to show only the densest parts of the flow. \label{fig:3d} }
\end{figure}

We performed two test runs with the initial conditions described in Section~\ref{Ini_shock_sph} with the parameters $\epsilon=0.0005, \lambda^{\rm eq}=3.65, a=0$ in full 3D, and with the resolution $N_{\rm r}$ x $N_{\theta}$ x $N_{\phi}$ equal to
384 x 192 x 64, and  256 x 128 x 96. The simulations were performed on the \textit{Cray} supercomupter cluster, typically using 16 nodes, and the message-passing interface supplemented with the hyperthreading technique was used, similarly as in \cite{2017ApJ...837...39J}.
The code is supposed to conserve the axi-symmetry of the initial state, which we confirm. Because no non-axisymmetric modes appeared, the solution is the same in each $\phi$-slice during the evolution (see Fig. \ref{fig:3d}).

\begin{figure}	
\begin{center}
 \includegraphics[width=0.5\textwidth]{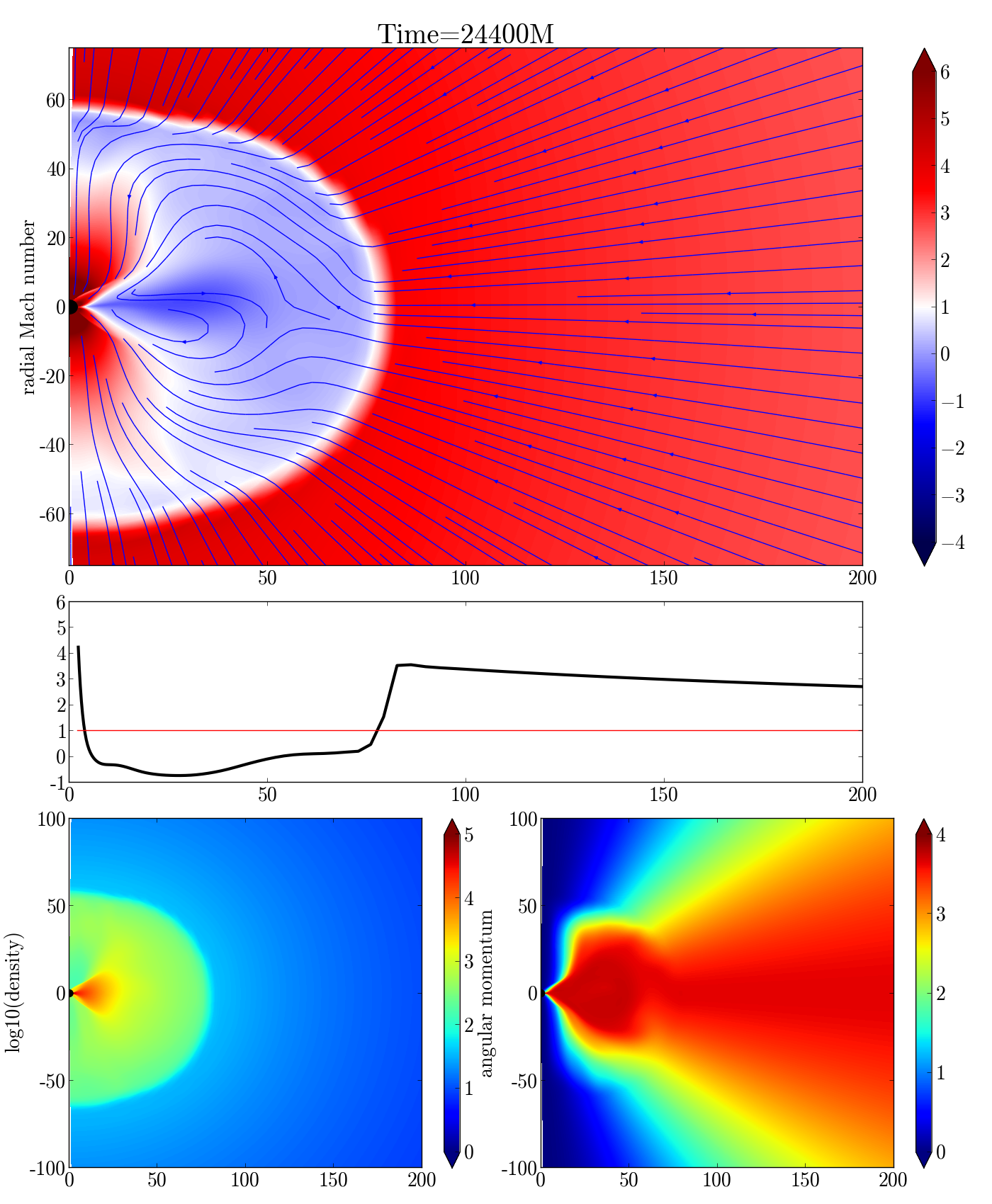}
 \caption{ Model \mIa{}: 3D simulation ($\epsilon=0.0005,\lambda^{\rm eq}=3.65,a=0.0$) with oscillating shock.  \label{K500} }
 \end{center}
\end{figure}

\begin{figure}
\begin{center}
 \includegraphics[width=0.5\textwidth]{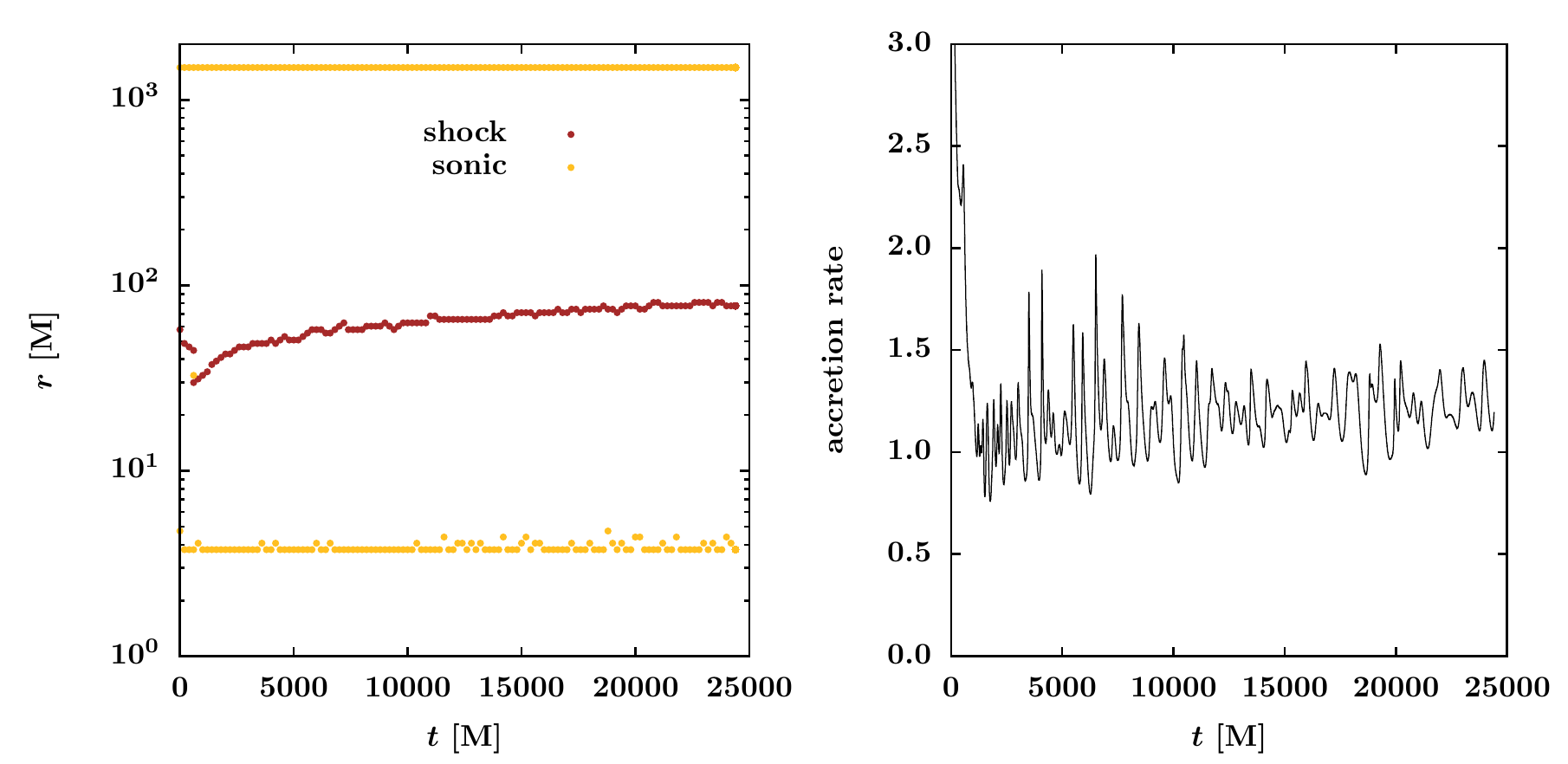}
 \end{center}
 \caption{  Model \mIa{}: 3D simulation ($\epsilon=0.0005,\lambda^{\rm eq}=3.65,a=0.0$) with oscillating shock. \label{K500_rs} }
\end{figure}

We were able evolved the system only up to $t_f = 34100$M, and $t_f=24400$M, for these two runs respectively, which is significantly shorter then in the 2D case. Hence, we cannot directly compare the long term evolution of the flow. However, the main features of the shock bubble evolution, which we observed in 2D case, appears also in 3D simulation. Mainly, the shock bubble has similar shape, as can be seen in Fig~\ref{K500}, and we observe similar oscillations of the bubble and of the mass accretion rate (Fig.~\ref{K500_rs}).

\section{Conclusions}\label{s:Conclusions}
In this paper we presented an extensive numerical study of the pseudo-spherical accretion flows with low angular momentum. The simulations were performed in the general relativistic framework on the Kerr background metric with the GR MHD code \texttt{HARMPI} in one, two and three dimensions.

As a first step, we provided set of 1D computations, which we compared to our earlier results obtained within the pseudo-Newtonian approach with the code ZEUS and published in \citet{our_paper}.
We confirm the qualitative properties of those simulations, which are especially the oscillation of the shock front for higher values of angular momentum and the possibility of the hysteresis effect, when the parameters of the flow are changing in time.

The oscillations of the shock front are connected with small oscillations of the value of angular momentum downward from the shock front, hence, they may be triggered by numerical errors at the shock front.
The amplitude of the oscillations reaches the maximum for intermediate shock positions and ceases again for very high shock position/angular momentum.

If the oscillations are physical, their frequency is in good agreement with the observed low frequency quasi-periodic oscillations seen in several black hole binary systems.
 Those are shifting in the range from hundreds of mHz up to few tens of Hz on the time scale of weeks (e.g. GRS1915+105 \citep{1999ApJ...513L..37M}, XTE J1550-564 \citep{1999ApJ...512L..43C}, GRO J1655-40 \citep{1999ApJ...522..397R} or GX 339-4 \citep{2012A&A...542A..56N}).
For a fiducial mass of the microquasar equal to 10M$_\odot$ this corresponds to the frequencies between $10^{-6}$M$^{-1}$ up to $10^{-3}$M$^{-1}$ in geometrized units\footnote{The time unit $1M$ equals to the time, which light needs to travel the half of the Schwarzschild radius of a black hole of mass M, hence for one solar mass $[t]=1M_\odot=\frac{GM_\odot}{c^3}=4.9255\cdot 10^{-6}s$.}, which is the same range as our observed values (see Fig.~\ref{1D_shock}, bottom panel for 1D results and Table~\ref{t:2D-models} for 2D simulations).

Hence, the change of the observed frequency during the onset and decline of the outburst is possibly connected with the change of angular momentum of the incoming matter (alternative explanation gives e.g. \cite{0004-637X-798-1-57}, who consider rather the change of viscosity parameter $\alpha$).
Such change can be either periodic and connected with the orbital motion of the companion, or caused by different conditions during the release of the matter. \cite{0004-637X-565-2-1161} proposed the scenario of the outburst of XTE J1550-564, in which the low angular momentum component of the accretion flow is released from the magnetic trap inside the Roche lobe of the black hole, kept by the magnetic field of the active companion. Depending on the distribution of angular momentum inside the trap and on the mechanism of breaking the magnetic confinement the angular momentum of the incoming matter from this region can slightly vary with time. Later the angular momentum of the low angular component could be affected by the interaction with the slowly inward propagating Keplerian disc.

To simulate such scenario, we have studied the case, when the angular momentum of the incoming matter changes periodically with time. As was shown on Fig.~\ref{1D_loop}, the behaviour of the shock front depends on whether the value of angular momentum crosses one or both of the values $\lambda_{\tt min}^{cr},\lambda_{\tt max}^{cr}$.
When the respective boundary is not crossed, the shock is moving slowly through the permitted region and its speed is determined by the parameters of the perturbation of the angular momentum.
When the angular momentum is in the corresponding range, the oscillations of the shock front develops.
When one of the the critical values is reached, the abrupt change of the flow geometry happens, while the flow transforms from one type of solution into another.
This transformation is achieved via the shock propagation, which is either accreted or expanding.

The velocity of the shock front agrees within the order of magnitude with the sound speed of the preshock medium, when the angular momentum is just slightly higher than $\lambda_{\tt max}^{cr}$. If the angular momentum is considerably higher, the formation and propagation of the shock can be even higher than the sound speed in the postshock medium (see Figs.~\ref{2D_Rs_position} and \ref{2D_Rs_position_eps}).
However, in nature we expect the first case to realize, because such situation can appear only if the angular momentum is increasing in the flow, which is in the Bondi-like configuration, hence it crosses smoothly through $\lambda_{\tt max}^{cr}$.

When extending our results to higher dimensions, the freedom of the dependence of angular momentum on the angle $\theta$ arises.
We have chosen two different configurations, described in Section \ref{Ini_shock}.
The comparison of the corresponding models 1D+sph and 1D+cone in Table~\ref{t:2D-models} shows, that in the latter case, the shock is placed at larger distances.
That can be understood as the consequence of the fact, that thanks to the relations given by Eqs. (\ref{uphi}) and (\ref{lambda-cone}) larger amount of matter posses the maximal value of angular momentum.
However, the main features of the solution, including the existence of the range of angular momentum enabling the long term shock presence, the shape of the resulting shock bubble and its oscillations, are similar in both cases.
The biggest difference between the two configurations is that in case of constant angular momentum in a cone, the peaks in mass accretion rate are not so prominent as in the case of angular momentum scaled by $\sin^2\theta$.
Hence, we conclude, that the physical processes in the low angular momentum accretion flows do not qualitatively depend on the exact distribution of angular momentum, but the observable consequences (e.g. the presence of prominent peaks and their amplitude) may be influenced by the geometry.

Similar conclusion holds also for the change of spin of the black hole. For all three considered values of spin of the black hole ($0.3$, $0.8$ and $0.95$) we have found a range of values of angular momentum of falling matter in which the long lasting oscillations of shock surface was observed. We have also observed, that the shock existence interval of angular momentum depends on the spin of the black hole: the higher the spin value, the lower both limiting angular momentum values between which the oscillating shock exists and the narrower is the corresponding interval.
Therefore, for rapidly spinning black holes, even a small change of the angular momentum of the incoming matter leads to significant changes in the flow itself (the existence, position and oscillations of the shock front) and also in the timing properties of the outgoing radiation (which we assume to be related to the accretion rate). Abrupt emergence, expansion or accretion of the shock which is connected with the crossing of the boundary of the shock existence region in parameter space and which leads to significant changes in the accretion rate, are more likely in the accretion flows around rapidly spinning black holes.

Simulations which we performed cover large variety of configurations: we considered different values of the spin of the central black hole, energy and angular momentum of the accreted gas, and even the distribution of the angular momentum of the matter.
Keeping all the other variables constant we have found the range of angular momentum in which there exist  oscillating shock solutions in all scenarios under consideration.
Hence the oscillating regime seems to be intrinsic to the low angular momentum accretion flows.
This finding is supported by \cite{2010MNRAS.403..516G} who also found an oscillating shock front in their hydrodynamical simulations in the pseudo-Newtonian framework.

Our simulations in two and three dimensions show, that for such parameters the oscillating shock front is long-lasting. That is an important ingredient for the POS model to be able to explain the QPO frequency change during outbursts of microquasars. 
However, the duration of our simulations corresponds to several tens of second for typical microquasar with $M=10  M_\odot$, which is still short in comparison with the time scale of the QPOs frequency change (weeks).
Moreover, we did not address this question from the point of view of an analytical stability analysis. 
Such analysis was provided by \cite{1980ApJ...235.1038M} for the spherical accretion onto non-rotating black hole and quite recently by \cite{BOLLIMPALLI2017153} for low angular momentum flow with standing shocks, who also report the stability of  the solution.

The dependence of the shock existence interval and consequently the position of the shock on the rotation of the black hole could be the probe   of the black hole spin.
However, there is a degeneracy between the spin and the angular momentum of the accreting matter, which itself is mostly unknown and hard to measure.
Hence, the constraints on the spin from the oscillating shock front model explaining QPOs can be posed only when there will be available better observations of the innermost accretion region or better models predicting the angular momentum of the LAF component.



\section*{Acknowledgements}

We acknowledge support from Interdisciplinary Center for Computational Modeling of the Warsaw University (grant GB66-3) and Polish National Science Center (2012/05/E/ST9/03914). PS is supported from Grant No. GACR-17-06962Y.

\bibliography{Sukova}

\label{lastpage}
\end{document}